\documentclass[preprint,12pt]{elsarticle}

\usepackage{tabularx}
\usepackage{amssymb}
\usepackage{svg}
\usepackage{amsmath}
\usepackage{amsthm}
\usepackage{amssymb}
\usepackage{float}
\usepackage{soul}
\usepackage{lscape}
\usepackage{hyperref}
\usepackage{cancel}
\usepackage{footnote}
\makesavenoteenv{tabular}
\makesavenoteenv{table}

\usepackage[ruled,linesnumbered]{algorithm2e}
\usepackage{booktabs}
\usepackage{multirow}
\usepackage{color}
\definecolor{blue}{rgb}{0,0,1}

\usepackage[normalem]{ulem} 
\usepackage{nomencl}
\makenomenclature

\begin{document}

\begin{frontmatter}

\title{HiFAKES: Synthetic High-Frequency NILM Data for NILM Models Diagnostics and Generalization Testing}

\author[inst1,inst2]{Ilia Kamyshev}

\author[inst1,inst2]{Sahar Moghimian Hoosh}

\author[inst1]{Henni Ouerdane}

\affiliation[inst1]{department={Center for Digital Engineering,}, organization={Skolkovo Institute of Science and Technology},
            addressline={30 Bolshoy Boulevard}, 
            city={Moscow},
            postcode={121205},
            country={Russia}}

\affiliation[inst2]{organization={Monisensa Development LLC},
            city={Moscow},
            country={Russia}}

\begin{abstract}
Monitoring electricity consumption at the appliance level is crucial for increasing energy efficiency in residential and commercial buildings. Using a single meter, the non-intrusive load monitoring (NILM) breaks down household consumption down to appliance-level, providing comprehensive insights into end-user electricity behavior. NILM models are typically trained on a household's total power consumption paired with submetered appliance labels. When sampled at high frequencies ($\geq$1 kHz), these datasets capture the full waveform characteristics, which significantly improves disaggregation accuracy and model generalization. Nevertheless, such datasets are scarce, collected from a limited number of households, and rarely include labels for power estimation, which complicates their use for model training, evaluation, or debugging. To overcome these limitations, we propose HiFAKES, a pre-trained synthetic data generator that can instantly generate unlimited amounts of fully labeled high-frequency NILM data, including aggregated and submetered current signatures. The data is ready-to-use and annotated for both load identification (classification) and power estimation (regression). Moreover, it allows simulating both seen and completely unseen scenarios of appliances' behavior with full control over parameters such as the number of appliance classes, operational modes, class similarity, brand diversity, and the number of concurrently running devices. Using HiFAKES, we propose a structured methodology to test the generalization of NILM models on simulated households with novel appliances and brands. The reliability of the synthetic data generated by HiFAKES is assessed using a domain-agnostic 3-dimensional metric. Results reveal that the generated signatures achieve high realism (93\% authenticity), closely resemble real-world data (84\% fidelity), and include a reasonable portion of novel, unseen signatures (5\%). 
\end{abstract}


\end{frontmatter}

\makenomenclature
\renewcommand{\nomgroup}[1]{%
  \ifthenelse{\equal{#1}{A}}{\item[\textbf{Abbreviations}]}{%
  \ifthenelse{\equal{#1}{S}}{\item[\textbf{Symbols}]}{%
  \ifthenelse{\equal{#1}{G}}{\item[\textbf{Greek Letters}]}{%
  \ifthenelse{\equal{#1}{U}}{\item[\textbf{Subscripts and Superscripts}]}{}}}}}

\nomenclature[G01]{\(\alpha\)}{Quantile threshold for real data support region}
\nomenclature[G02]{\(\beta\)}{Quantile threshold for synthetic data support region}
\nomenclature[G03]{\(\epsilon_{\text{sep}}\)}{Class separability parameter}
\nomenclature[G04]{\(\mu_k\)}{Centroid of the \(k\)-th Gaussian blob (cluster)}
\nomenclature[G05]{\(\tau\)}{Relative size of the training set}
\nomenclature[G06]{\(r\)}{Pearson correlation coefficient}
\nomenclature[G07]{\(\sigma_k\)}{Spread parameter controlling the width of cluster \(k\)}
\nomenclature[G08]{\(\sigma^2_{g(i)}\)}{Explained variance by the \(i\)-th principal component of synthetic data}
\nomenclature[G09]{\(\sigma^2_{r(i)}\)}{Explained variance by the \(i\)-th principal component of real data}
\nomenclature[G10]{\(\Sigma_k\)}{Covariance matrix of the \(k\)-th Gaussian blob}
\nomenclature[G11]{\(\theta\)}{Phase angle between the current signature and the grid voltage}

\nomenclature[S01]{\(A\)}{Number of aggregate signals}
\nomenclature[S02]{\(\mathbf{A}\)}{Binary activation matrix}
\nomenclature[S03]{\(B\)}{Total number of brands}
\nomenclature[S04]{\(B(x, r)\)}{Euclidean ball centered at \(x\) with radius \(r\)}
\nomenclature[S05]{\(C\)}{Total number of clusters}
\nomenclature[S06]{\(D\)}{Number of appliance classes}
\nomenclature[S07]{\(E\)}{Reconstruction error}
\nomenclature[S08]{\(\mathcal{D}\)}{Generated dataset}
\nomenclature[S09]{\(L\)}{Number of principal components used}
\nomenclature[S10]{\(M\)}{Number of operational modes}
\nomenclature[S11]{\(N\)}{Number of synthetic samples}
\nomenclature[S12]{\(o_g\)}{Center of hypersphere for synthetic data}
\nomenclature[S13]{\(o_r\)}{Center of hypersphere for real data}
\nomenclature[S14]{\(P_a\)}{Target power shares matrix}
\nomenclature[S15]{\(P_{\alpha}\)}{$\alpha$-precision metric}
\nomenclature[S16]{\(\Phi(\cdot)\)}{Projection function used for computing $\alpha$- and $\beta$-supports}
\nomenclature[S17]{\(R_{\beta}\)}{$\beta$-recall metric}
\nomenclature[S18]{\(r_{\alpha}\)}{Radius of \(\alpha\)-support region}
\nomenclature[S19]{\(r_{\beta}\)}{Radius of \(\beta\)-support region}
\nomenclature[S20]{\(S\)}{Apparent power}
\nomenclature[S21]{\(S_g^{\beta}\)}{\(\beta\)-support region for synthetic data}
\nomenclature[S22]{\(S_r^{\alpha}\)}{\(\alpha\)-support region for real data}
\nomenclature[S23]{\(T\)}{Samples per signature}
\nomenclature[S24]{\(v\)}{Grid voltage waveform}
\nomenclature[S25]{\(W\)}{Active power}
\nomenclature[S26]{\(\mathbf{W}_g^\top\)}{PCA projection matrix for synthetic data}
\nomenclature[S27]{\(\mathbf{W}_r\)}{PCA reconstruction matrix for real data}
\nomenclature[S28]{\(X_a\)}{Synthetic aggregated current signatures matrix}
\nomenclature[S29]{\(X_g\)}{Synthetic submetered current signatures matrix}
\nomenclature[S30]{\(X_r\)}{Real submetered current signatures matrix}
\nomenclature[S31]{\(Z_g\)}{PCA (latent) representation of synthetic data}
\nomenclature[S32]{\(Z_r\)}{PCA (latent) representation of real data}
\nomenclature[S33]{\(\tilde{x}_{g,\beta}^*\)}{Nearest synthetic sample in \(\beta\)-support}
\nomenclature[S34]{\(\tilde{x}_{r}^*\)}{Nearest real sample}
\nomenclature[S35]{\(y_g\)}{Generated cluster IDs}
\nomenclature[S36]{\(y_{\text{brand}}\)}{Appliance brand IDs}
\nomenclature[S37]{\(y_{\text{class}}\)}{Appliance class IDs}

\nomenclature[U01]{\(g\)}{Synthetic/generated signal or matrix }
\nomenclature[U02]{\(r\)}{Real signal or matrix}
\nomenclature[U03]{\(a\)}{Aggregate signal}

\nomenclature[A01]{AC}{Alternating Current}
\nomenclature[A02]{AOT}{Admittance Over Time}
\nomenclature[A03]{AUC-ROC}{Area Under the Receiver Operating Characteristic Curve}
\nomenclature[A04]{CPU}{Central Processing Unit}
\nomenclature[A05]{FF}{Form Factor}
\nomenclature[A06]{FFT}{Fast Fourier Transform}
\nomenclature[A07]{FITPS}{Frequency-Invariant Transformation of Periodic Signals}
\nomenclature[A08]{GAN}{Generative Adversarial Network}
\nomenclature[A09]{GBM}{Gaussian Blob Modeling}
\nomenclature[A10]{HiFAKES}{High-Frequency Synthetic Appliance Signatures Generator}
\nomenclature[A11]{ICA}{Independent Component Analysis}
\nomenclature[A12]{KNN}{K-Nearest Neighbors}
\nomenclature[A13]{MAE}{Mean Absolute Error}
\nomenclature[A14]{ML}{Machine Learning}
\nomenclature[A15]{NILM}{Non-Intrusive Load Monitoring}
\nomenclature[A16]{NMF}{Non-Negative Matrix Factorization}
\nomenclature[A17]{PCA}{Principal Component Analysis}
\nomenclature[A18]{PIASG}{Physics-Informed Appliance Signature Generator}
\nomenclature[A19]{RMS}{Root Mean Square}
\nomenclature[A20]{SVD}{Singular Value Decomposition}
\nomenclature[A21]{TC}{Temporal Centroid}
\nomenclature[A22]{VIT}{Voltage-Current Trajectory}
\nomenclature[A23]{WE}{Wavelet Energy}
\nomenclature[A24]{XGBoost}{Extreme Gradient Boosting}
\nomenclature[A25]{$R^{2}$}{Coefficient of Determination}

\printnomenclature

\section{Introduction}
\label{sec:sample1}
Global electricity consumption is growing rapidly due to the ongoing electrification, a growing number of data centers, and increased usage of electrical devices, specifically cooling systems \cite{strielkowski2021renewable}. According to a recent IEA report, the world's electricity consumption is expected to increase at its highest rate in recent years, reaching over 4\% annually until 2027. This is equivalent to adding more than Japan's yearly electricity consumption each year from now until 2027 ~\cite{iea2}. The rapid growth of electricity demand across the world is a major cause of environmental problems, and decreases power supply security \cite{locmelis2020benchmarking}. This requires ambitious policy initiatives, such as implementing demand-side efficiency measures \cite{barbar2023impact}. 

The residential sector alone is a major contributor to the growing demand with almost 25-30\% of the final energy consumption in the European Union~\cite{umbark2020energy}. That is why households are considered among the primary targets for adopting energy-saving programs policies. This aligns with EU's zero-emission building targets for 2028, which mandate the residential sector to implement energy-efficient solutions \cite{IEA_2024, savaresi2016paris}. Improving the energy efficiency of the existing residential buildings through renovations is often costly and time-consuming. As such, low-cost alternatives such as increasing user awareness, encouraging energy-saving behaviors, and installing energy-efficient appliances seem more practicable for residential consumers \cite{ruano2019nilm, YOUSSEF2023127793}.

In this context, non-intrusive load monitoring (NILM), also known as energy disaggregation, is a promising solution on the end-user side that can be seamlessly integrated into existing smart meters \cite{BECKEL2014397}. NILM provides detailed insights into the energy consumption of household appliances \cite{pereira2018performance} by using machine- or deep-learning techniques to estimate the state and energy consumption of individual appliances from the total consumption of a household using a single metering point \cite{Hart1992NonintrusiveAL}. This technology enables appliance-level, and real-time feedback on electricity consumption to consumers, grid operators, utility companies, automated building controllers, and appliance manufacturers \cite{schirmer2022non}. Studies indicate that such fine grained insights can help end-users reduce their electricity consumption by 10–15\% \cite{Event_detection,CARRIEARMEL2013213}.

To improve NILM performance, numerous models have been developed, ranging from classic machine learning algorithms to advanced deep learning architectures \cite{azad2023non, kelly2015neural}, including convolutional neural networks \cite{shin2019subtask, chen2019scale}, convolutional sequence to sequence models \cite{chen2018convolutional}, recurrent neural networks  \cite{de2021recurrent}, long short-term memory networks \cite{hwang2022nonintrusive}, and transformer-based models \cite{kamyshev2025coldconcurrentloadsdisaggregator, varanasi2024stnilm}.

Despite advances in modeling techniques and the growing number of publicly available datasets, implementing NILM in practice still faces limitations, especially regarding the generalizability of models \cite{d2019transfer, kaselimi2022towards}. In fact, models tend to perform well during training and testing on seen data, but their accuracy drops significantly when transferred to new homes with different appliance types, and unseen brands \cite{ahmed2020edge}.
The generalizability of NILM models is in fact affected by dataset size and quality, appliance variety (type and brand), and more importantly the quality of extracted features \cite{d2019transfer}. Depending on the sampling rate of the signal, different types and resolutions of features can be extracted \cite{tabanelli2021trimming}, with sampling rates above 1 kHz often containing detailed features of appliances' electrical behaviors. These microscopic features include spectral envelopes, transient power, shape characteristics, and transient voltage-current trajectories \cite{iqbal2021critical} which are used to distinguish overlapped appliances with similar behaviors \cite{klemenjak2019electricity, wu2020nonintrusive, chavan2022iedl}.  

While high sampling rate data can significantly improve the performance and generalization of NILM models \cite{da2021deepdfml}, research in this area remains limited compared to low-frequency NILM, mainly due to challenges in data availability and data handling.
There exist publicly available high-sampling-rate NILM datasets such as UKDALE \cite{kelly2015uk}, BLOND~\cite{Kriechbaumer2018BLONDAB}, WHITED~\cite{inproceedings}, COOLL~\cite{picon2016cooll}, PLAID~\cite{plaid}, BLUED~\cite{filip2011blued}, and REDD \cite{kolter2011redd}, recently reviewed in \cite{papageorgiou2025nilm}. 
However, none of these datasets is fully labeled for the task of power estimation in NILM (i.e., multi-label regression) \cite{silva2024review}. In more details, certain datasets such as BLOND, BLUED, PLAID, LIT \cite{renaux2020dataset}, and LILAC \cite{kahl2019measurement}, include event-based labels (ON/OFF events), but they do not have true labels on the individual contribution of each appliance to the aggregate signal. Some datasets such as UK-DALE have both aggregate and submetered power measurements but they are not synchronized and not ready for straightforward use by a NILM model. While manual labeling should be done before further processing, this task is challenging and error-prone due to inconsistencies between aggregated and submetered signals within the datasets \cite{nour2023data}. In addition, these datasets often suffer from missing data and noise, requiring time-consuming and costly preprocessing steps, which in turn affect the generalizability of NILM models \cite{may2010data}. Note that most existing high-frequency datasets are collected from only a small number of households (typically not more than five) \cite{iqbal2021critical}, and often contain a limited variety of appliance combinations, which is far from the millions of other households. Hence, data-related challenges are precluding progress in high-frequency NILM research, which needs to be addressed \cite{shin2019data}.

Synthetic data generation and data augmentation are promising solutions to manage these limitations \cite{geng2025diffusion}. They are well-established methods in other areas, e.g., natural language processing to handle the data scarcity problem, establish a rich testing environment, and enhance model generalization \cite{goyal2024systematic}.
\textit{Data augmentation}, which should not be confused with \textit{fully synthetic data}, does not introduce new classes but modifies existing classes of real-world data to generate new variants. On the other hand, fully synthetic data is created by generative models that can learn the distribution of the real-world data, and then generate entirely new samples even with unseen classes. The synthetic data preserves the essential characteristics of real-world signals without the need for timely and costly data collection \cite{goyal2024systematic}. More importantly, the generated data comes with exact labels, which, in the context of NILM, are appliance states for load identification and usage shares for power estimation. This makes synthetic data a valuable tool for creating baselines, supporting model developments, and validating algorithms, especially in rare or completely new scenarios. Importantly, it should be noted that synthetic data is not a replacement for real data, but by using it, the performance of NILM models can be evaluated over a variety of testing and debugging experiments.

For low-frequency NILM, there exist several studies on data augmentation models \cite{geng2025diffusion,7778841, 8340657,antgen, SynD, donnal2022nilm, li2022energy}. The first diffusion-based model to produce realistic, multi-state, and low-noise load data for low-frequency NILM was recently proposed by Geng et al. \cite{geng2025diffusion}. Only a small number of studies, nevertheless, have made an effort to create generative models for high-frequency NILM, with the majority using augmentation techniques \cite{nour2023data, HENRIET2018268, weisshaar2020expansion, held2019generation}. While data augmentation methods like those employed by Henriet et al. \cite{HENRIET2018268}, Wei{\ss}haar et al. \cite{weisshaar2020expansion}, and Held et al. \cite{held2019generation} can expand existing datasets, they still require considerable amounts of real-world measurements as a starting point for training. In contrast, approaches like Physics-Informed Appliance Signatures Generator (PIASG) \cite{kamyshev2023physics}, represent a significant advancement by generating completely novel signatures without requiring any training data. The common limitations of all existing generative models for high-frequency NILM are diversity and flexibility. In fact, none of them can simulate:
i) novel brands i.e., a different model or manufacturer within the same appliance class;
ii) novel operational regimes (e.g., a washing machine’s known mode that was not recorded in training data); iii) fully-labeled training and testing aggregate signatures including fractions of power consumed by each active appliance; iv) signatures in a fully controlled manner i.e., class similarity, class diversity, number of classes, number of brands per class, number of regimes per class, number of simultaneously working appliances. Additionally, except for our prior work (PIASG), none of these methods can generate novel appliance classes, i.e., appliance types not seen in the training data (e.g., introducing a dishwasher when only washing machines were present).

Note that simulating all these cases amounts to emulating what would happen in a completely different residential setting which is the major reason for using synthetic datasets. Without a customizable data generation tool with tunable parameters, one cannot evaluate how their models will perform across different homes, and NILM debugging will remain slow and not reproducible.

\subsection*{Contributions}
To address these limitations, we introduce HiFAKES, which is a pre-trained synthetic data generator designed to automatically create an arbitrary, unlimited number of realistic high-frequency appliance signatures and aggregate signatures. HiFAKES is based on the valuable finding from real-world data that each appliance class and its brand-level variations form a distinct cluster in a compressed feature space. We construct this space using principal component analysis (PCA), allowing for efficient modeling of appliance signatures as dense clusters. Unlike with other generative models, a HiFAKES user can fully control key properties of the generated dataset. The pre-trained model can generate each dataset in a few seconds which includes submetered signatures, fully labeled aggregated traces, and scenarios with different difficulty levels. Our primary contributions are summarized as follows:
\begin{itemize}
    \item \textbf{HiFAKES} produces fully labeled appliance-level and aggregate high-frequency signals (up to 30 kHz) with fine-grained control over appliance properties. The generated data is of standard machine learning format (input and target) and requires no further processing, thus enabling the rapid debugging of high-frequency NILM models as well as testing their generalization potential.
    
    \item We propose a structured methodology for \textbf{controlled generalization testing} in NILM using HiFAKES. Our framework isolates three critical real-world challenges: class separability (appliances similarity), simultaneous appliance operation (background load), and brands diversity. This allows for systematic testing of NILM models across these dimensions, and choosing the most robust model to all these factors.
   
    \item We enable a fully customizable data generation down to the appliance level, meaning it can create datasets based on a user's requirements, such as the total number of signatures, appliance classes (or types), number of brands per class, operational regimes, and number of simultaneously working appliances, as well as the class separability and intra-class diversity.

    \item Given the increasing number of generative models in NILM research, standardized evaluation metrics are required to determine how representative and reliable the synthetic datasets are. We do this by using a 3-dimensional, domain-agnostic metric that can evaluate and measure the generalization, diversity, and fidelity of any generative model. We recommend using this metric as a standardized reference in future NILM studies.
    
\end{itemize}

The remainder of this paper is structured as follows. Section~\ref{literature} reviews the existing synthetic data generation and augmentation techniques in NILM with a focus on high-frequency datasets. In Section~\ref{sec:method}, we present the HiFAKES framework, its design assumptions, and modular architecture for generating synthetic submetered and aggregate signals. In section~\ref{sec:eval}, we comprehensively evaluate the fidelity, diversity, and authenticity of the generated signatures. In the same section, we apply the 3D-metric test for the first time in the context of NILM generative models which can be regarded as a standard for assessing the reliability of synthetic data in future research. 
A case study is presented in Section~\ref{sec:case}, where we show how HiFAKES can be used in practice to test the generalization capabilities of NILM models under various conditions, including brand diversity, class separability, and simultaneous appliance operation. In Section~\ref{sec:dis}, HiFAKES is compared with previous synthetic data generators. We also discuss the advantages of using PCA for latent space modeling,  limitations of HiFAKES, and suggest directions for future work. 

\section{Literature review}
\label{literature}
Data generation models are widely used to enrich datasets when the available data is limited, imbalanced, unlabeled or does not represent diverse practical scenarios. This can be done either by using data augmentation techniques, which typically increase the amount and diversity of existing data, or through data synthesizers, which generate as much data as needed from scratch by learning the distribution (e.g., shape and variance) and structure (e.g., feature correlations) of the original dataset \cite{hernandez2022synthetic}.  For NILM, generative models are mainly based on the augmentation techniques appropriate for time-series data, such as jittering (noise addition), scaling, warping (of magnitude, time, or frequency), resampling, and signal combination \cite{iwana2021time}. More advanced techniques include deep generative models such as generative adversarial networks (GANs), progressively growing GAN, and EEG-GAN \cite{liu2024non}.
For instance, TraceGAN \cite{tracegan} uses a conditional 1D Wasserstein GAN, trained on the low-frequency REFIT dataset (1/8 Hz), to generate synthetic power traces. However, in this work, we do not intend to compare our approach with low-frequency data augmentation models such as SmartSim~\cite{7778841}, AMBAL~\cite{8340657}, ANTgen \cite{antgen}, SynD~\cite{SynD}, and NILM-Synth~\cite{donnal2022nilm}, as their scope differ substantially.

The number of generative models for high-frequency disaggregation data remains limited. As shown in Table \ref{table1:comp}, this research direction has only recently begun to develop, with a few notable works that merit detailed discussion. The SHED dataset \cite{HENRIET2018268} is among the first models that generates augmented high-frequency disaggregation data for commercial buildings. It decomposes appliance electrical current signatures into a predetermined number of components using a semi-non-negative matrix factorization algorithm. This number is chosen to ensure that the signal-to-noise ratio between the actual signature and the residual (the difference between the real and reconstructed signatures) exceeds 50 dB, indicating high-quality reconstruction. In fact, the synthetic signatures are generated by adding Gaussian noise to the reconstructed signatures, and then the total building consumption is computed by summing the synthetic appliance signatures along with modeled device activations and background noise. 
Likewise, synthetic aggregated data was generated by summing current consumption measurements from individual appliances, using the frequency-invariant transformation for periodic signals (FIT-PS) \cite{held2019generation}. Using real measurements, FIT-PS was applied as a synchronization technique to align the phase and frequency of all signals \cite{held2019generation}. This enabled a consistent combination of individual recordings, even when they come from different datasets or have different sampling rates. Later, the FIT-PS technique was also used in  \cite{weisshaar2020expansion}, where a cycle expansion method was introduced to construct longer-duration appliance traces from short switching cycles. These extended traces were then superimposed to generate aggregate signals with control over number of overlapping devices, and timing offsets.

In \cite{nour2023data}, each appliance signals were reconstructed as a combination of harmonics scaled by an exponential envelope. By swapping stable parameters (e.g., harmonic amplitudes), they generated up to $n^2$ synthetic signals from $n$ measurements, which were aggregated into synthetic multi-appliance signals. However, the method assumed exponential decay, which cannot accurately model non-exponential or complex transients, and the algorithm required detailed inputs. In a recent work, a hybrid deep generative model, TimeGAN combined with FIT-PS was proposed to synthesize both aggregated and submetered high-frequency NILM data \cite{xiao2025non}. 

Compared to all previously mentioned works, PIASG \cite{kamyshev2023physics} can be considered as the first generative model capable of synthesizing truly unseen appliance signatures with no need for real-world measurements. In \cite{kamyshev2023physics} two physics-informed data generators were proposed: one for high-sampling-rate and another for low-sampling-rate signals, respectively. High-frequency signatures were generated using a mathematical model that incorporates characteristics such as exponential decay of harmonics, phase changes between $-\pi/2$ and $\pi/2$ radians, variations in the spectrum of AC cycles over time, and exponential decay of transient processes. PIASG involves randomly sampling sets of parameters that control these characteristics, with each set corresponding to a unique synthetic appliance. These parameters are then used as location parameters for predefined statistical distributions (e.g., normal, half-normal). From these distributions, parameters for a predefined number of appliance signatures are sampled. The sampled parameters are then substituted into the universal mathematical model for an appliance to produce authentic appliance signatures. This approach offers significant advantages including transparent and intuitive control over the underlying distributions, the ability to simulate appliances without requiring input data or a training process, and flexibility in generating signatures at different sampling frequencies. However, it lacks the capability to generate unseen brands because of the shared physical model, aggregate signatures and class separability.

\begin{landscape}
\begin{table}[htbp]
\caption{Comparison of high-frequency NILM data augmentation and synthetic data generation methods}
\renewcommand{\arraystretch}{1.5}
\centering
\resizebox{\linewidth}{!}{%
\begin{tabular}{|c|c|c|c|c|c|c|c|c|c|c|c|}
\hline
Ref & Year & 
\multicolumn{2}{c|}{\begin{tabular}[c]{@{}c@{}}Generated\\ signal type\footnote{Agg = Aggregate signals; Sub = Submetered signals.}\end{tabular}} & 
Modeling & 
\begin{tabular}[c]{@{}c@{}}Augments\\ known\\ classes\end{tabular} &
\begin{tabular}[c]{@{}c@{}}Generates\\ unseen\\ classes\end{tabular} &
\begin{tabular}[c]{@{}c@{}}Generates\\ unseen\\ brands\end{tabular} &
\begin{tabular}[c]{@{}c@{}}Unseen\\ combinations\end{tabular} &
\begin{tabular}[c]{@{}c@{}}Adjustable\\ parameters\footnote{Indicates whether the method allows user modification of generation parameters.}\end{tabular} &
\multicolumn{2}{c|}{\begin{tabular}[c]{@{}c@{}}Label type\footnote{LI = load identification (classification); PE = power estimation (regression).}\end{tabular}} \\
\cline{3-4} \cline{11-12}
& & Agg & Sub & & & & & & & LI & PE \\
\hline
\cite{HENRIET2018268} & 2018 & $\checkmark$ & $\checkmark$ & 
\begin{tabular}[c]{@{}c@{}}Semi-NMF decomposition +\\ Gaussian noise injection\end{tabular} & 
$\checkmark$ & $\times$ & $\times$ & $\checkmark$ & $\times$ & \textbf{$\checkmark$} & $\times$ \\
\hline
\cite{held2019generation} & 2019 & $\checkmark$ & $\times$ & 
\begin{tabular}[c]{@{}c@{}}FIT-PS + element-wise\\ summation of real signals\end{tabular} & 
$\checkmark$ & $\times$ & $\times$ & $\checkmark$ & $\times$ & $\checkmark$ & $\times$ \\
\hline
\cite{weisshaar2020expansion} & 2020 & $\checkmark$ & $\times$ & 
\begin{tabular}[c]{@{}c@{}}FIT-PS + steady-state\\ expansion + summation\end{tabular} & 
$\times$ & $\times$ & $\times$ & $\checkmark$ & $\times$ & \textbf{$\checkmark$} & $\times$ \\
\hline
\cite{nour2023data} & 2023 & $\times$ & $\checkmark$ & 
\begin{tabular}[c]{@{}c@{}}Harmonic envelope\\ modeling + decay\end{tabular} & 
$\checkmark$ & $\times$ & $\times$ & $\times$ & $\times$ & \textbf{$\checkmark$} & $\times$ \\
\hline
\cite{kamyshev2023physics} & 2023 & $\times$ & $\checkmark$ & 
\begin{tabular}[c]{@{}c@{}}Physics-based signal\\ generation\end{tabular} & 
$\checkmark$ & $\checkmark$ & $\times$ & $\times$ & $\checkmark$ & \textbf{$\checkmark$} & $\times$ \\
\hline
\cite{xiao2025non} & 2025 & $\checkmark$ & $\checkmark$ & 
FIT-PS + deep generative modeling & 
$\checkmark$ & $\times$ & $\times$ & $\checkmark$ & $\times$ & $\checkmark$ & $\times$ \\
\hline

\textbf{Ours} & \textbf{2025} & \textbf{$\checkmark$} & \textbf{$\checkmark$} & 
\textbf{\begin{tabular}[c]{@{}c@{}}Latent space sampling via\\ non-isotropic Gaussian modeling\end{tabular}} & 
\textbf{$\checkmark$} & \textbf{$\checkmark$} & \textbf{$\checkmark$} & \textbf{$\checkmark$} & \textbf{$\checkmark$} & \textbf{$\checkmark$} & \textbf{$\checkmark$} \\

\hline
\end{tabular}%
}
\label{table1:comp}
\end{table}
\end{landscape}

\section{Methodology}
\label{sec:method}
\subsection{Overview of HiFAKES}

HiFAKES is the first unsupervised tool for instantaneous generation of custom, high-frequency, and fully labeled datasets for designing unseen scenarios in NILM research. This generative model requires only unlabeled submetered appliance signatures to be trained. The complete training of HiFAKES takes 45 seconds using the Intel Core i9-11900K CPU. Once trained, it can produce \textit{any required number} of fully-labeled aggregated signatures. For instance, 100,000 signatures are generated in approximately one minute. In \href{https://github.com/arx7ti/hifakes}{the source-code repository}, we uploaded the pre-trained version of HiFAKES so that the end-user can utilize the tool immediately for model debugging and developments.

\begin{figure}[t]
  \centering
  \includegraphics[width=0.75\textwidth]{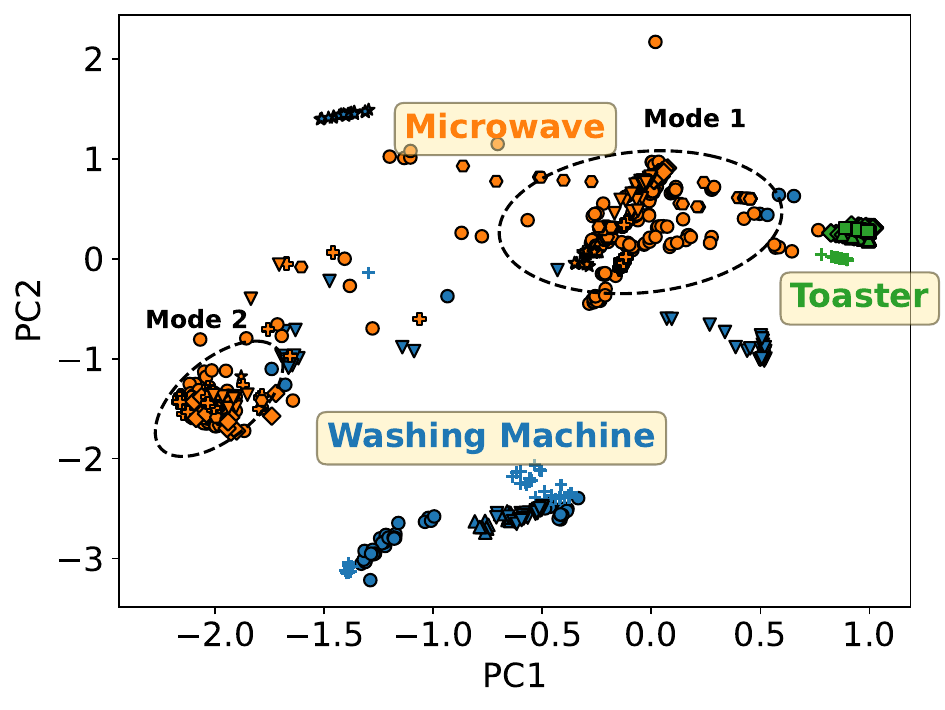}
  \caption{Two-dimensional PCA projection of  high-frequency current signatures from the PLAID and WHITED datasets for three appliance types: washing machine, microwave, and toaster. Different types of appliances naturally form distinct clusters in the latent space. The microwave cluster splits into two sub-clusters (mode 1 and mode 2), indicating different operational modes. Different marker shapes within each cluster represent different manufacturing brands of the same appliance type.}
  \label{fig:auto-clusters}
\end{figure}

HiFAKES combines all the essential features of its predecessor models within a unified framework: augmenting existing signatures of appliances, creating unseen classes of appliances, and generating signatures for simultaneous operation of any number of appliances. More importantly, it introduces key and novel contributions, including the ability to generate \textit{different brands} and distinct \textit{operational modes} of a given appliance, as well as the level of appliances similarity. This is achieved by modeling high-frequency signatures in a latent space --- a compressed representation where the dimensionality is significantly lower than that of the original time-domain data. In fact, our approach is based on the observation that in a latent space e.g., obtained using PCA, appliances of the same type (e.g., heaters), even from different brands, exhibit more similar signatures to each other than to those of entirely different appliance types (e.g., washing machines). In fact, the latent space organizes high-frequency signatures into distinct clusters of appliances as illustrated in Fig.~\ref{fig:auto-clusters}. This data structure enables the use of Gaussian blob modeling (GBM) which is a well-established and effective method for simulating clustered data \cite{guyon2003design}, by invoking the assumption that clusters are sampled using a multivariate Gaussian distribution.

\begin{figure}[ht!]
  \centering
  \includegraphics[width=1\textwidth]{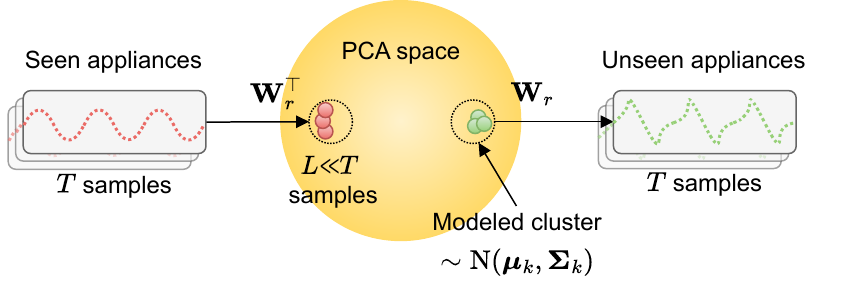}
  \caption{Illustration of high-frequency signature generation using Gaussian blobs modeling in PCA space and a reconstruction matrix. \( T \) is the number of samples in the original time-domain signature and \( L \) is the number of samples in the compressed space required to accurately represent the original signature. \( \mathbf{W}_r^\top \) is the transformation matrix used to map signatures from the time domain to PCA space, and \( \mathbf{W}_r \) is the reconstruction matrix to map latent signatures back to time-domain. \( \boldsymbol{\mu}_k \) denotes the cluster centroid, and \( \boldsymbol{\Sigma}_k \) is the random covariance matrix.}
  \label{fig:high-level}
\end{figure}

\begin{figure}[ht!]
  \centering
  \includegraphics[width=1\textwidth]{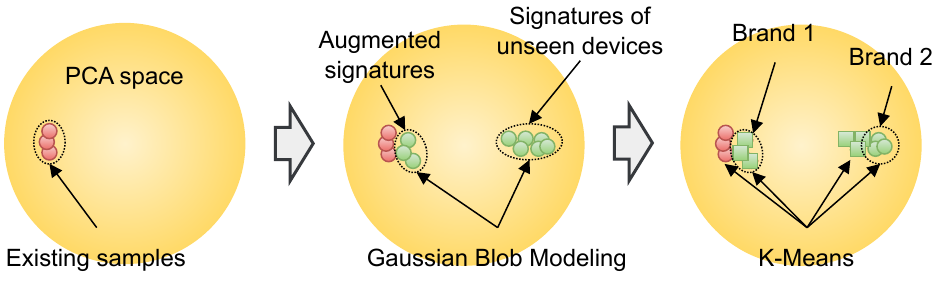}
  \caption{Illustration of the step-by-step process for generating signatures in the PCA space. The left panel demonstrates the distribution of existing high-frequency appliance signatures in the PCA space. The middle panel illustrates the result of Gaussian blob generation with two randomly sampled clusters, some of which overlap with existing data (augmentation) while others are positioned further away (novel classes simulation). The right panel demonstrates the annotation of brand labels by using K-means clustering: squares stand for brand 1 while circles for brand 2.}
  \label{fig:brandsmodeling}
\end{figure}

Thus, in HiFAKES, we generate different appliances using GBM, where each cluster corresponds to an operational mode, and a group of clusters forms an appliance class (type). Samples within each cluster correspond to current signatures. Then, we apply a K-means clustering approach within each cluster to divide it into sub-clusters representing different brands of a specific appliance type. To obtain time-domain signatures we compute the matrix product between the latent signatures and the inversed transformation matrix obtained by PCA. A high-level overview of the process of generating appliance signatures in the compressed PCA space is shown in Fig.~\ref{fig:high-level} and brands modeling is shown in Fig.~\ref{fig:brandsmodeling}.

HiFAKES produces signatures with a high level of realism and diversity by removing the need for explicit modeling of high-frequency signatures in either the time or the frequency domain. Instead, it models a distribution of signatures in the compressed latent space. This approach has been applied in other fields such as, e.g., geometrical machine learning or manifold learning, and it automatically enables a wider coverage of signatures representations in terms of seen and unseen appliances. Seen or known signatures are covered when the modeled distribution overlaps with the real one, otherwise unseen signatures occur. HiFAKES leverages linear methods such as PCA to effectively compress the high-resolution current signatures, with 0.07 A of mean absolute error compared to the original waveforms and 99\% of the original variance is preserved. For HiFAKES, we set 50 principal components for the latent space which a represents a compression rate of 600 by comparison with the 30,000 samples in the time domain when working with either the PLAID or WHITED training datasets.

We utilize PCA instead of other methods such as independent component analysis (ICA) or non-negative matrix factorization (NMF) due to its simplicity, computational efficiency, interpretability, and sufficient reconstruction accuracy. PCA produces well-separated clusters of appliances that approximate Gaussian shapes, which significantly simplifies latent distribution modeling. Additionally, PCA provides an orthogonal latent space, implying that the components are uncorrelated and eliminating the need for explicit covariance modeling. ICA was not adopted in this work due to its high sensitivity to noise, especially in scenarios where appliance signatures exhibit only small variations around a typical ``average signature'' (see Fig.~\ref{fig:auto-clusters}). Under such conditions, ICA’s reliance on higher-order statistical independence may lead to unstable or inconsistent reconstructions. As for NMF and its variants, they generate non-orthogonal and often correlated bases, unlike PCA. This introduces the added complexity of modeling inter-feature covariances in the latent space.

\subsection{Assumptions}
HiFAKES is based on the following minor assumptions to significantly enhance the computational performance:
(i) the voltage waveform is purely sinusoidal; (ii) all synthetic signatures are generated within the same power grid; (iii) each AC cycle maintains a fixed frequency, indicating no frequency fluctuations; (iv) compressed signatures form non-isotropic Gaussian clusters; (v) each cluster of signatures corresponds to an operational mode of an appliance.

Assumption (i) is standard in most electrical engineering applications. Assumption (ii) upholds Kirchhoff's current law, meaning that the signature which represents the total current at a common node equals the sum of the signatures of the individual currents flowing into that node. Assumption (iii) is introduced because, over a short observation window (e.g., 1 minute), changes in frequency are negligibly small. This assumption enables a matrix-based representation of the data, which is essential for the efficient execution of the generative model, especially when constructing synthetic aggregate signatures from submetered data. Assumptions (iv) and (v) are motivated by empirical observations of the distribution of compressed high-frequency signatures in PCA-transformed space, as illustrated in Fig.~\ref{fig:auto-clusters}. While we do not claim that appliance classes and their operational modes inherently form Gaussian clusters in all latent representations, our results show that PCA produces a space in which the signatures exhibit well-separated, compact groupings. These groupings are sufficiently regular to be approximated by a mixture of multivariate Gaussian distributions for the purpose of generative modeling. These assumptions are rather the empirically justified approximation that enables efficient synthesis of realistic signatures within the HiFAKES framework.

\SetKwComment{Comment}{/* }{ */}

\subsection{Framework}

\begin{figure}[ht!]
  \centering
  \includegraphics[width=.7\textwidth]{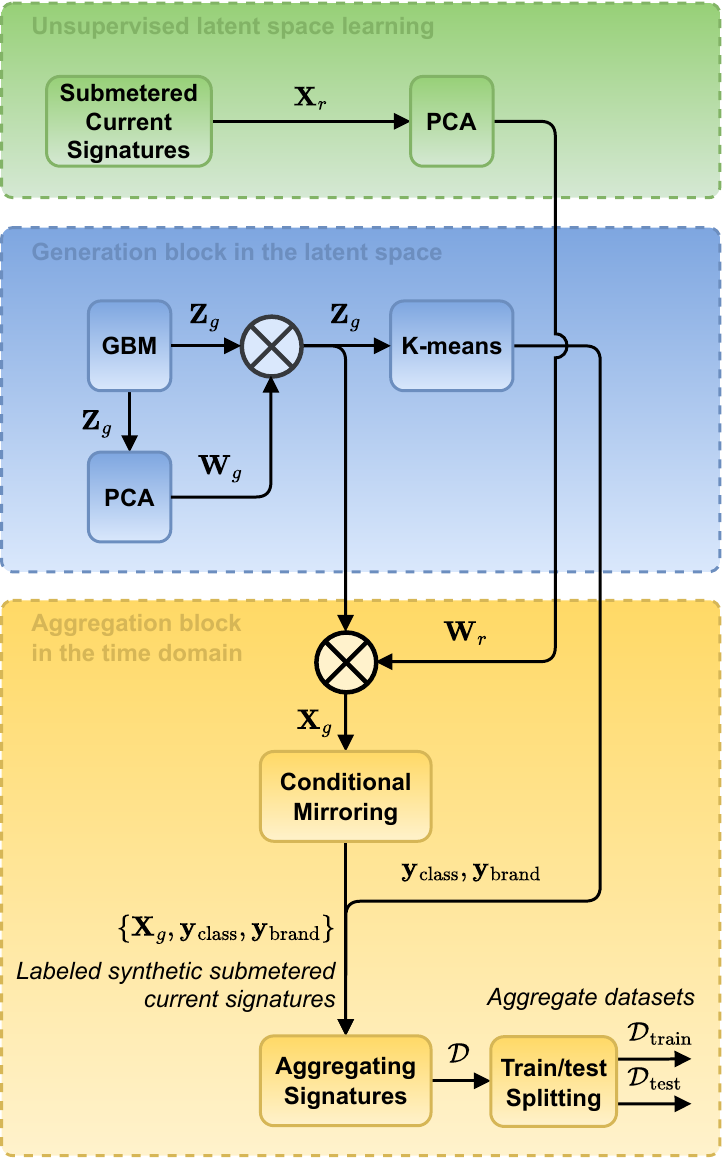}
  \caption{Diagrammatic representation of HiFAKES framework.}
  
  \label{fig:framework}
\end{figure}

Based on the overview above, the framework contains three main down-stream architectural blocks (see Fig.~\ref{fig:framework}):

\begin{enumerate}
    \item\textbf{Unsupervised latent space learning} applying PCA to \textit{real submetered signatures} to obtain the reconstruction matrix $\mathbf{W}_r$ that will be further used for reconstructing (decoding) latent synthetic signatures.
    \item\textbf{Generation block in the latent space} to model the latent distribution of appliances and their operational modes using GBM and to annotate brands.
    \item\textbf{Aggregation block in the time domain} to reconstruct synthetic time-domain signatures from the latent space; to simulate simultaneous operation of appliances; to assign target power shares; and to divide the obtained dataset into train and test subsets.
\end{enumerate}

\subsubsection{Unsupervised latent space learning}
To learn a compact and meaningful representation of appliance signatures, HiFAKES employs PCA to extract latent features from high-dimensional ($T=30,000$ samples) real-world submetered electrical signatures. Specifically, we decompose the observed waveforms into $L=50$ principal components which results in a mean absolute reconstruction error of approximately 0.07 A, and 99\% of variance preserved. The number $L=50$ was chosen based on the cumulative explained variance threshold.

We use two publicly available high-frequency NILM datasets as a source of real-world submetered signatures: PLAID~\cite{plaid} and WHITED~\cite{inproceedings}. These are among the few datasets that provide appliance-level current and voltage waveforms sampled at tens of kilohertz capturing both transient and steady-state electrical behaviors.

\begin{figure*}[ht!]
  \centering
  \includegraphics[width=1\textwidth]{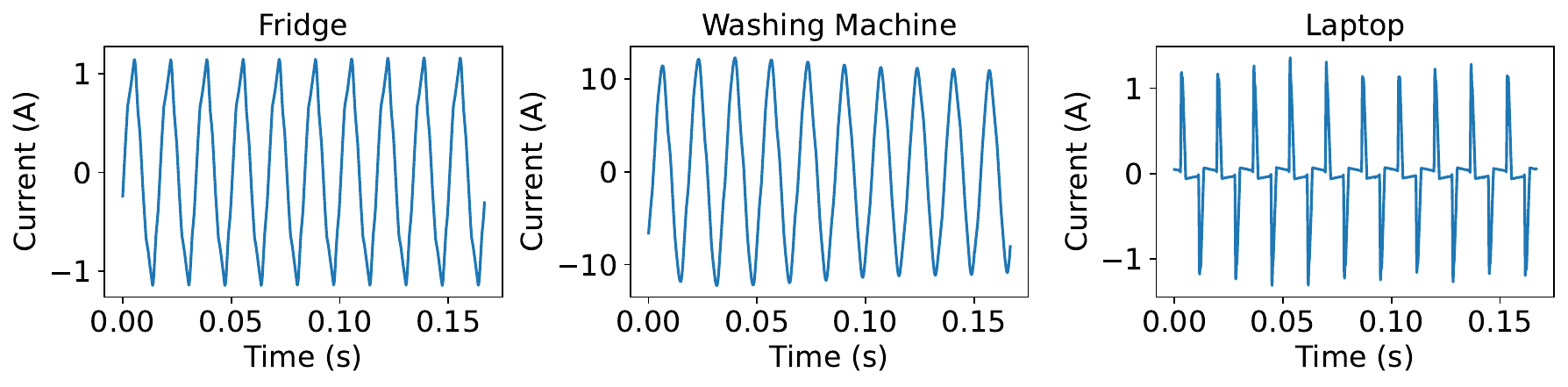}
  \caption{Examples of current signatures from real measurements of a fridge, washing machine, and laptop. A 0.15-s slice (from a 1-s waveform) is shown to emphasize distinct shape and amplitude features.}
  \label{fig:signatures}
\end{figure*}

All training waveforms are first unified to a common sampling rate of 30 kHz through resampling to standardize temporal resolution across datasets. We then segment each waveform into operational phases using a Bayesian change point detection method by isolating transitions such as start-up events, mode switches, and steady-state intervals.
Each segment is then cropped into fixed-length windows of $N = 30,000$ samples, representing one second of measurements. To ensure consistency in the fundamental harmonic across all segments, we apply the FITPS algorithm~\cite{held2018frequency}, which eliminates the fluctuations in the mains frequency and makes all signatures mains-frequency-invariant i.e., independent of the value of the frequency of 50 or 60 Hz, which here has become irrelevant. This preprocessing step ensures compliance with assumption (iii) of HiFAKES and guarantees the voltage-current phase alignment during the aggregation stage of the framework. These one-second traces form a library of appliance-level signatures (see Fig.~\ref{fig:signatures}) that are used for obtaining the reconstruction matrix $\textbf{W}_r$ via PCA.

As previously mentioned, PCA is a linear dimensionality reduction technique that extracts uncorrelated components from multivariate data. It projects the data onto a lower-dimensional subspace defined by orthogonal directions that capture the maximum variance. Given a mean-centered data matrix, \( \mathbf{X}_r \in \mathbb{R}^{N \times T} \), where each row is an \( T \)-dimensional observation and there are \( N \) such samples, PCA seeks a reconstruction matrix \( \mathbf{W}_r \in \mathbb{R}^{L \times T} \), with \( L \leq T \), such that the compressed latent representation \( \mathbf{Z}_r \in \mathbb{R}^{N \times L} \) captures the directions of maximal variance:
\begin{equation}
    \mathbf{Z}_r = \mathbf{X}_r \mathbf{W}_r^\top
\end{equation}
where $\textbf{W}_r^\top\in\mathbb{R}^{T\times L}$ is a projection matrix, and index $r$ stands for real data. The original data can then be approximately reconstructed using a linear projection from the latent space:
\begin{equation}
\label{eq:inverse_pca}
    \mathbf{X}_r \approx \mathbf{Z}_r \mathbf{W}_r
\end{equation}

In this work, we quantify reconstruction accuracy using the mean absolute error (MAE) between the original and reconstructed data:
\begin{equation}
    E = \frac{1}{TL} \sum_{i=1}^N \sum_{j=1}^T \left| \mathbf{X}_r^{(i,j)} - [\mathbf{Z}_r \mathbf{W}_r]^{(i,j)} \right|
\end{equation}

\noindent With PCA, we compute the matrix \( \mathbf{W}_r \) by performing singular value decomposition (SVD) on \( \mathbf{X}_r \), a mean-centered matrix in \(\mathbb{R}^{N \times T} \):
\begin{equation}
    \mathbf{X}_r = \mathbf{U} \boldsymbol{\Sigma} \mathbf{V}^\top
\end{equation}
where \( \mathbf{U} \in \mathbb{R}^{N \times N} \) contains the left singular vectors, \( \boldsymbol{\Sigma} \in \mathbb{R}^{N \times K} \) is a diagonal matrix of singular values, and \( \mathbf{V} \in \mathbb{R}^{T \times T} \) contains the right singular vectors. The reconstruction matrix \( \mathbf{W}_r \in \mathbb{R}^{L \times T} \) is formed by selecting the top \( L \) rows of \( \mathbf{V}^\top \), corresponding to the \( L \) leading principal components:
\begin{equation}
    \mathbf{W}_r = \mathbf{V}_{1:L}^\top
\end{equation}
Each row of \( \mathbf{W}_r \) represents a principal direction, sorted according to the magnitude of explained variance of the original data under the constraint of orthogonality.

\subsubsection{Generation block in the latent space}
HiFAKES starts generating synthetic appliance signatures in the PCA space. As mentioned above, one advantage of PCA space is that it is orthogonal, which guarantees that there is no correlation between latent components. This eliminates the necessity of explicit modeling of covariances between latent features as it would be required in case of NMF. Another key advantage of working in the compressed space is the significant reduction of generation time. In our case, for $L=50$ latent features per signature the theoretical reduction of time spent for generating a single signature is at least 1,000 times lower compared to time-domain, where each signature contains $T=30,000$ samples.

Taking into account assumptions (iv) and the considerations above, one practical approach to simulate synthetic signatures in the latent PCA space is to first generate non-isotropic Gaussian blobs. Non-isotropicity follows from the fact that the clusters are not necessarily symmetric but can be skewed or stretched in space across different dimensions. In this work, and in line with assumption (v), we interpret each dense cluster in PCA space as a distinct operational mode of an appliance. Since real appliances often operate in multiple modes (e.g., idle, heating, standby), each synthetic appliance class is modeled as a collection of such clusters. HiFAKES allows the user to specify the number of operational modes per appliance class using the parameter \( M \). Hence, the total number of clusters to be generated is $C=D\cdot M$, where $D$ is the number of appliances' classes specified by the user.

Each Gaussian blob represents a localized point cloud in an $L$-dimensional space whose points are sampled from a multi-variate Gaussian distribution defined by a centroid \( \boldsymbol{\mu}_k \) of size $L$ and a random covariance matrix \( \boldsymbol{\Sigma}_k \) of size $L\times L$. Thus, a synthetic signature \( \mathbf{z}_g \in \mathbb{R}^L \) of $k$-th cluster is generated as:

\begin{equation}
\label{eq:blob}
\mathbf{z}_g \sim \text{N}(\boldsymbol{\mu}_k, \boldsymbol{\Sigma}_k)
\end{equation}

\noindent where the random covariance matrix reads:
\begin{equation}
    \boldsymbol{\Sigma}_k = \sigma_k^2\frac{1}{M} \cdot \frac{1}{\operatorname{Tr}(\mathbf{S})} \mathbf{S}, \quad
    \text{with} \quad \mathbf{S} = \mathbf{P} \mathbf{P}^\top
\end{equation}
and the elements of the matrix $\mathbf{P}\in\mathbb{R}^{L\times L}$ are sampled from a standard normal distribution. Since for real-life data, not all clusters have the same density, it is fair to introduce a random spread of clusters by multiplying all elements of $\boldsymbol{\Sigma}$ by $\sigma_k^2$, where $\sigma_k\sim\text{U}(\sigma_{\min}, \sigma_{\max})$, and the parameters of $\text{U}$ are user-defined.

To generate $C$ clusters, their respective centroids are sampled $C$ times from a uniform distribution $\boldsymbol{\mu}_k\sim\text{U}(\textbf{z}_{\text{min}},\textbf{z}_{\text{max}})$. Each such centroid represents a location of a blob in the PCA space. Coordinates of locations are bounded by $\textbf{z}_{min}$ and $\textbf{z}_{max}$ computed from the PCA space of real signatures.

In addition to controlling cluster diversity with \( \sigma_k \), we explicitly regulate inter-class separation using the parameter \( \varepsilon_{\text{sep}} \). Specifically, we rescale and center the randomly initialized centroids \( \boldsymbol{\mu}_k \) as:
\begin{equation}
    \boldsymbol{\mu}_k \gets \varepsilon_{\text{sep}} \cdot (2\boldsymbol{\mu}_k - 1)
\end{equation}
Here, \( \leftarrow \) denotes assignment. This transformation ensures that all centroids $\mathbf{M}\in\mathbb{R}^{C\times C}$ are symmetrically distributed around the origin and spaced proportionally to \( \varepsilon_{\text{sep}} \). By reducing \( \varepsilon_{\text{sep}} \), appliance classes are forced closer together in the latent space, increasing their similarity in the time-domain and making NILM tasks more challenging. Conversely, larger values create more separable, easily distinguishable classes of appliances.

\begin{algorithm}[ht!]
\DontPrintSemicolon
\SetNoFillComment
\SetSideCommentLeft
\caption{\texttt{GBM} function for generating clustered data}
\label{alg:GBM}
\KwData{$N$, $L$, $D$, $M$, $\varepsilon_{\text{sep}}$, $z_{\min}$, $z_{\max}$, $\sigma_{\min}$, $\sigma_{\max}$}
\KwResult{$\mathbf{Z}_g$, $\mathbf{y}_g$}

\Comment{Compute number of clusters and components}
$C \gets D \cdot M$\;

\Comment{Sample cluster centers and separate them}
$\mathbf{M} \gets \texttt{uniform}([z_{\min},z_{\max}], \text{size} = [C, L])$\;
$\mathbf{M} \gets \varepsilon_{\text{sep}}\cdot (2 \mathbf{M} - 1)$\;

\Comment{Distribute samples across clusters}
$\mathbf{n}_{\text{dist}} \gets \left\lfloor \frac{N}{C} \right\rfloor \cdot \mathbf{1}$\;

\Comment{Initialize data and labels}
$\mathbf{Z}_g \gets \text{zeros}(N, L)$\;
$\mathbf{y}_g \gets \text{zeros}(N)$\;

$a \gets 0$\;

\For{$k \gets 0$ \KwTo $C - 1$}{
    $b \gets a + \mathbf{n}_{\text{dist}}[k]$\;

    \Comment{Generate random semi-definite covariance matrix}
    $\sigma_k\gets \texttt{uniform}(\sigma_{\min}, \sigma_{\max})$\;
    $\mathbf{P} \gets \texttt{normal}(0, 1, \text{size} = [L, L])$\;
    $\mathbf{S} \gets \mathbf{P} \cdot \mathbf{P}^\top$\;
    $\boldsymbol{\Sigma}_k \gets \sigma_k^2\frac{1}{M}\frac{\mathbf{S}}{\operatorname{Tr}(\mathbf{S})}$\;
$\boldsymbol{\mu}_k\gets\mathbf{M}[k]$\;
    \Comment{Generate Gaussian cluster}
    $\mathbf{Z}_g[a{:}b] \gets \texttt{multivariateNormal}(\boldsymbol{\mu}_k, \mathbf{\Sigma}_k, \text{size} = \mathbf{n}_{\text{dist}}[k])$\;
    $\mathbf{y}_g[a{:}b] \gets k$\;

    $a \gets b$\;
}
\end{algorithm}

Some generated clusters may partially overlap with real clusters (see Fig.~\ref{fig:pcaspace}) in the PCA space. In this case, the generated data \textit{augment} existing classes of appliances or, in other words, creates \textit{authentic copies} of existing signatures of an appliance. In contrast, non-overlapping clusters are treated as entirely new (unseen) appliances, enabling HiFAKES to simulate novel appliances beyond those present in the original submetered dataset.

Although the synthetic signatures are initially generated using GBM in an abstract Euclidean space, they do not inherently conform to the structure of the real PCA space. Specifically, uncorrelated axes and a variance profile are not aligned with real principal components. To enforce this structure, we first apply an auxiliary PCA to the synthetic data, yielding a projection matrix \( \mathbf{W}_g^\top \), and transform the latent signatures accordingly:
\begin{equation}
    \mathbf{Z}_g \leftarrow \mathbf{Z}_g \mathbf{W}_g^\top
\end{equation}

To further match the scale of each principal component, we normalize each column $i$ of \( \mathbf{Z}_g \) by its corresponding standard deviation \( \sigma_g^{(i)} \) (from the synthetic PCA) and rescale it using the standard deviation \( \sigma_r^{(i)} \) of the matching component from the real data:
\begin{equation}
    \mathbf{Z}_g^{(i)} \leftarrow \mathbf{Z}_g^{(i)} \cdot \frac{\sigma_r^{(i)}}{\sigma_g^{(i)}}
\end{equation}

\noindent This two-step transformation ensures that the synthetic latent space exhibits both the decorrelated axes and the variance profile of the real PCA space. Thus, allowing for further decoding of synthetic latent signatures into the time domain.

Once this alignment is achieved, we can use the synthetic PCA space to define appliance-specific structures. To introduce brand-level diversity, we further subdivide each operational mode (i.e., each blob) into \( B \) segments, each representing a different brand. This partitioning is performed using the K-means clustering algorithm, which is computationally efficient and relies on the Euclidean distance in the PCA space to assign points to compact and distinct groups. Each resulting segment is labeled with a brand identification label \( \mathbf{y}_{\text{brand}} \), corresponding to the K-means cluster index.

Class labels \( \mathbf{y}_{\text{class}} \) are derived by enumerating all generated clusters and grouping them based on their parent appliance class. Specifically, if \( \mathbf{y}_g \) denotes the cluster index assigned to each point, class labels are computed as:
\begin{equation}
    \mathbf{y}_{\text{class}} \leftarrow \left\{ \left\lfloor \frac{k}{M} \right\rfloor : k \in \mathbf{y}_g \right\}
\end{equation}

The full procedure for generating $N$ synthetic submetered appliance signatures of $D$ classes of appliances and $B$ brands is detailed in Algorithm~\ref{alg:makeSubmetered}.

\subsection{Aggregation block in the time-domain}

The next step is to decode the compressed representations of signatures into the time domain. The previously obtained reconstruction matrix $\mathbf{W}_r$ are substituted into Eq.~\ref{eq:inverse_pca} to get the time-domain high-resolution synthetic current signatures $\mathbf{X}_g$. Some reconstructed signatures may appear in counter-phase with the grid voltage due to their location in previously unseen regions of the latent space. To correct this, we apply a conditional mirroring procedure (see Algorithm~\ref{alg:condMirror}), which flips the time-series signal if its phase angle \( \theta \) violates the physical constraint \( -\pi/2 \leq \theta \leq \pi/2 \). The mirroring operation for a reconstructed current signature $\mathbf{x}\in\mathbf{X}_g$ is defined as:
\begin{equation}
    \textbf{x} \leftarrow -1\cdot\textbf{x},
\end{equation}
i.e., the signal is flipped over vertical axis. To detect whether mirroring is required, we compute the Pearson correlation coefficient \( r \) between the signature \( \textbf{x} \) and the reference grid voltage \( \textbf{v} \) (as per assumptions (i) and (ii)). The correlation is given by:
\begin{equation}
    r = \frac{\sum_{t=1}^{T}(x_t - \bar{x})(v_t - \bar{v})}{\sqrt{\sum_{t=1}^{T}(x_t - \bar{x})^2} \cdot \sqrt{\sum_{t=1}^{T}(v_t - \bar{v})^2}},
\end{equation}
where \( \bar{x} \) and \( \bar{v} \) are the mean values of the current and voltage signals, respectively. If \( r < 0 \), indicating the signals are out of phase, the signature is mirrored to restore physical consistency.

\begin{algorithm}[ht!]
\DontPrintSemicolon
\SetNoFillComment
\SetSideCommentLeft
\caption{\texttt{makeSubmetered} function for creating synthetic signatures in the PCA space}
\label{alg:makeSubmetered}
\KwData{$N$, $L$, $D$, $M$, $B$, $\varepsilon_{\text{sep}}$, $(z_{\min}, z_{\max})$, $(\sigma_{\min}, \sigma_{\max})$}
\KwResult{$\mathbf{Z}_g$, $\mathbf{y}_{\text{class}}$, $\mathbf{y}_{\text{brand}}$}

\Comment{Generate clustered data}
$(\mathbf{Z}_g, \mathbf{y}_g) \gets \texttt{GBM}(N, L, D, M, \varepsilon_{\text{sep}},z_{\min}, z_{\max}, \sigma_{\min}, \sigma_{\max})$\;

\Comment{Construct PCA space}
$(\mathbf{W}_g,\boldsymbol{\sigma}^2_g) \gets \texttt{PCA}(\mathbf{Z}_g, L)$\;
$\mathbf{Z}_g \gets \mathbf{Z}_g\mathbf{W}_g^\top$\;
$\mathbf{Z}_g \gets \mathbf{Z}_g / \boldsymbol{\sigma}_g\cdot\boldsymbol{\sigma}_r$\;

\Comment{Assign synthetic brand labels}
$\mathbf{y}_{\text{brand}} \gets \texttt{zeros}(\texttt{size}(\mathbf{y}_g))$\;

\ForEach{$k \in \texttt{unique}(\mathbf{y}_g)$}{
    \Comment{Partition each cluster}
    $\mathbf{p} \gets \texttt{KMeans}(B,\mathbf{Z}_g[\mathbf{y}_g=k])$\;
    $\mathbf{y}_{\text{brand}}[\mathbf{y}_g = k]=\mathbf{p}$\;
}
\Comment{Collapse mode-level labels to class-level}
$\mathbf{y}_{\text{class}} \gets \mathbf{y}_g / M$\;

\end{algorithm}

\begin{algorithm}[ht!]
\DontPrintSemicolon
\SetNoFillComment
\SetSideCommentLeft
\caption{\texttt{condMirror} function for waveform phase alignment}
\label{alg:condMirror}
\KwData{$\mathbf{X}_g$, $\mathbf{v}$}
\KwResult{$\mathbf{X}_g$}
$\mathcal{I}\gets\texttt{range}(\texttt{rows}(\mathbf{X}_g))$\;
\ForEach{$i\in\mathcal{I}$}{
$r\gets\texttt{sign}(\texttt{pearson}(\mathbf{X}_g^{(i)},\mathbf{v}))$\;
\If{$r<0$}{
   $\mathbf{X}_g^{(i)}\gets -1\cdot\mathbf{X}_g^{(i)}$\;
  }
}
\end{algorithm}

Next, the simultaneous operation of appliances is simulated by generating aggregate current consumption signatures that satisfy Kirchhoff's law under the given assumptions (i) and (ii). This is achieved by constructing a sparse activation $A\times N$ matrix \( \mathbf{A} \), where $A$ and $N$ are the total number of aggregate and submetered signatures, respectively. Each row of the matrix $\mathbf{A}$ specifies a linear combination of submetered signatures that contribute to the total consumption for a given scenario. The resulting aggregate signature is computed as:

\begin{equation}
    \mathbf{X}_a^{(i)} = \sum_{j=1}^{N} a_{i,j} \, \mathbf{X}_g^{(j)},
\end{equation}
where \( \mathbf{X}_g^{(j)} \) is a submetered signature of appliance \( j \), and \( a_{i,j} \in \mathbf{A} \) is a binary indicator such that \( a_{i,j} = 1 \) if appliance \( j \) is active for given scenario \( i \), and \( 0 \) otherwise.

To simulate realistic aggregate signatures, we generate scenarios where each scenario $i$ has $K_i \sim \text{U}(K_{\min}, K_{\max})$ simultaneously operating appliances with $K_{\min}$, $K_{\max}$ defining the concurrency bounds. The binary activation matrix $\mathbf{A}$ enforces that scenario $i$ activates exactly $K_i$ appliances through the constraint:

\begin{equation}
\sum\limits_{j=1}^{N} \mathbf{A}^{(i,j)} = K_i
\end{equation}

For the power estimation task, it is essential to calculate the target power shares of each device in the total consumption. We first define the matrix $\mathbf{P}_a\in\mathbb{R}^{A\times D}$ of target power shares, where each element $\mathbf{P}_a^{(i,d)}$ represents the fraction of total power in scenario \(i\) attributed to an appliance class \(d\):
\begin{equation}
    \mathbf{P}_a^{(i,d)} = \frac{1}{T}\sum_{t=1}^T v_t \sum_{j\in\mathcal{I}(i)}\mathbf{X}_g^{(j,t)},
\end{equation}
where $\mathcal{I}(i)$ is the index set of appliances that are active in scenario $i$ and belong to an appliance $d$, and $v_t$ is an instance of a grid's voltage signature $\mathbf{v}$. So, the quadruple: matrix of aggregate signatures $\mathbf{X}_a$, class labels $\mathbf{y}_{\text{class}}$, brand labels $\mathbf{y}_{\text{brand}}$, and target power shares $\mathbf{P}_a$, constitutes the dataset $\mathcal{D}$.

Finally, the dataset \( \mathcal{D} \) is split into training and testing subsets using one of two strategies: \emph{uniform-split} or \emph{brand-split}. In the uniform-split, a random subset of samples of a predefined share ($\tau$) is assigned to \( \mathcal{D}_{\text{train}} \), with the remaining samples assigned to \( \mathcal{D}_{\text{test}} \). In the brand-split, \( \mathcal{D}_{\text{train}} \) contains all samples whose brands belong to a randomly selected subset of brands, while the remaining samples are placed in \( \mathcal{D}_{\text{test}} \).

Thus, HiFAKES synthesizes $A$ aggregate signatures from a scalable set of $N$ submetered signatures, hierarchically organized into $D$ classes, $M$ modes per class, and $B$ brands per class. The aggregation block is detailed in Alg.~\ref{alg:makeAggregate} and the full pipeline of HiFAKES is listed in Alg.~\ref{alg:makeDatasets}.

\begin{algorithm}[ht!]
\DontPrintSemicolon
\SetNoFillComment
\SetSideCommentLeft
\caption{\texttt{makeAggregate} function for generating aggregate NILM dataset with the use of synthetic latent signatures}
\label{alg:makeAggregate}
\KwData{$A$, $K_{\min}$, $K_{\max}$, $\mathbf{Z}_g$, $\mathbf{W}_r$,$\mathbf{y}_{\text{class}}$, $\mathbf{y}_{\text{brand}}$, $\text{split\_mode}$, $\tau$}
\KwResult{$\mathcal{D}_{\text{train}}$, $\mathcal{D}_{\text{test}}$}
\Comment{Decoding latent signatures}
$\mathbf{X}_g\gets\mathbf{Z}_g\mathbf{W}_g$\;
\Comment{Post-processing}
$\mathbf{X}_g\gets\texttt{condMirror}(\mathbf{X}_g, \mathbf{v})$\;
\Comment{Initialize variables}
$N\gets\texttt{rows}(\mathbf{X}_g)$\;
$D\gets\texttt{size}(\texttt{unique}(\mathbf{y}_{\text{class}}))$\;
$B\gets\texttt{size}(\texttt{unique}(\mathbf{y}_{\text{brand}}))$\;
$\mathbf{A}\gets\texttt{zeros}(A,N)$\;
$\mathbf{Y}_{\text{class}}\gets\texttt{zeros}(A,D)$\;
$\mathbf{P}_a\gets\texttt{zeros}(A,D)$\;
$\mathbf{Y}_{\text{brand}}\gets\texttt{zeros}(A,B)$\;
$\mathcal{I}\gets\texttt{range}(\texttt{rows}(\mathbf{X}_g))$\;
\Comment{Aggregate}
\ForEach{$i\in\mathcal{I}$}{
$K\gets\texttt{randint}(K_{\min}, K_{\max})$\;
$\mathcal{J}\gets\texttt{randint}(0, N, \text{size}=K)$\;
\ForEach{$j\in\mathcal{J}$}{
 $\mathbf{A}^{(i,j)}\gets 1$\;
 $d\gets\mathbf{y}_{\text{class}}^{(i)}$\;
 $\mathbf{Y}_{\text{class}}^{(i,d)}\gets1$\;
 $\mathbf{P}_a^{(i,d)}\gets\texttt{mean}
 (\mathbf{X}_g^{(i,j)},\mathbf{v})$\;
 $b\gets\mathbf{y}_{\text{brand}}^{(i)}$\;
 $\mathbf{Y}_{\text{brand}}^{(i,b)}\gets1$\;
}
}

$\mathbf{X}_a\gets \mathbf{A}\mathbf{X}_g$\;
$\mathcal{D}\gets \{\mathbf{X}_a,\mathbf{Y}_{\text{class}},\mathbf{Y}_{\text{brand}},\mathbf{P}_a\}$\;
\Comment{Splitting}
$(\mathcal{D}_{\text{train}},\mathcal{D}_{\text{test}})\gets\texttt{train\_test\_split}(\mathcal{D},\ \text{train\_size}=\tau,\ \text{mode}=\text{split\_mode})$\;
\end{algorithm}

\begin{algorithm}[ht!]
\DontPrintSemicolon
\SetNoFillComment
\SetSideCommentLeft
\caption{\texttt{makeDatasets} function to generate fully labeled synthetic NILM datasets using HiFAKES framework}
\label{alg:makeDatasets}
\KwData{$A$, $K_{\min}$, $K_{\max}$, $N$, $L$, $N$, $D$, $M$, $B$, $\varepsilon_{\text{sep}}$, $(z_{\min}, z_{\max})$, $(\sigma_{\min}, \sigma_{\max})$, $\text{split\_mode}$, $\tau$,\ $\text{weights\_path}$}
\KwResult{$\mathcal{D}_{\text{train}}$,$\mathcal{D}_{\text{test}}$}
\eIf{\text{weights\_path}}{
    $\mathbf{W}_r\gets \texttt{load\_weights}(\text{weights\_path})$\;
    $L\gets\texttt{rows}(\mathbf{W}_r)$\;
}{
$(\mathbf{W}_r,\sigma_r)\gets\texttt{PCA}(\mathbf{X}_r, L)$\;
}
$(\mathbf{Z}_g,\mathbf{y}_{\text{class}},\mathbf{y}_{\text{brand}})\gets\texttt{makeSubmetered}(N, L, D, M, B, \varepsilon_{\text{sep}}, z_{\min}, z_{\max}, \sigma_{\min}, \sigma_{\max})$\;
$(\mathcal{D}_{\text{train}},\mathcal{D}_{\text{test}})\gets\texttt{makeAggregate}(A, K_{\min}, K_{\max}, \mathbf{Z}_g, \mathbf{W}_r,\mathbf{y}_{\text{class}}, \mathbf{y}_{\text{brand}}, \text{split\_mode},\ \tau)$\;
\end{algorithm}

\section{Evaluation}
\label{sec:eval}

To assess the quality of the generated signatures, we first conduct a visual comparison between generated and real signatures, followed by a quantitative evaluation using the 3D-metric test which comprises $\alpha$-precision, $\beta$-recall, and authenticity. 

\subsection{Visual Comparison}

\begin{figure}[ht!]
  \centering
  \includegraphics[width=1.\textwidth]{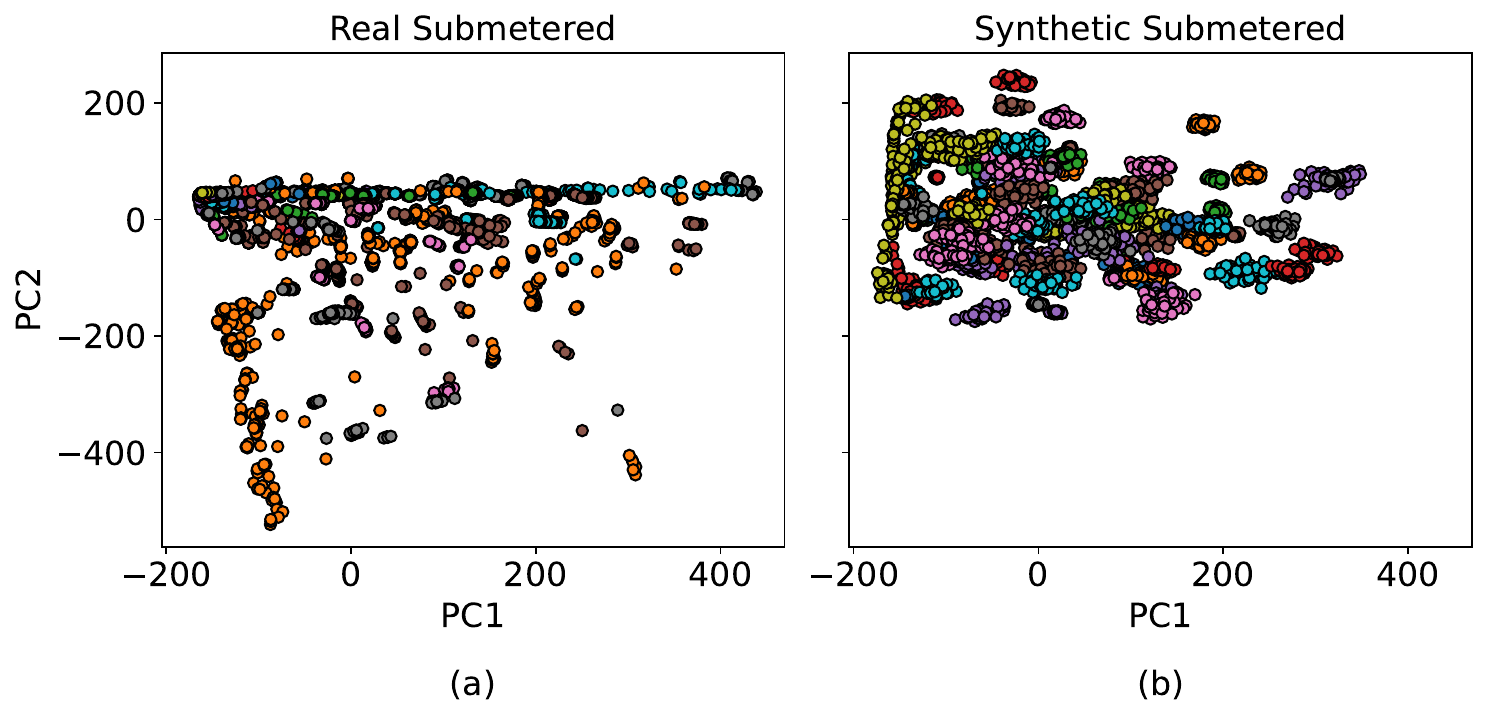}
  \caption{PCA projection of (a) real and (b) synthetic appliance signatures.
}
  \label{fig:pcaspace}
\end{figure}

For visual analysis of the generated data, we project both real ($\mathbf{X}_r$) and synthetic ($\mathbf{X}_g$) signatures into the PCA space using first two principal components. As shown in Fig.~\ref{fig:pcaspace}, the synthetic samples cover the PCA space in the directions of real data, indicating that, in some cases, HiFAKES augments existing appliances. On the other hand, once can notice certain samples reside outside of the hull of real samples. In this case, HiFAKES produces previously unseen appliances. 

\begin{figure}[ht!]
  \centering
  \includegraphics[width=1.\textwidth]{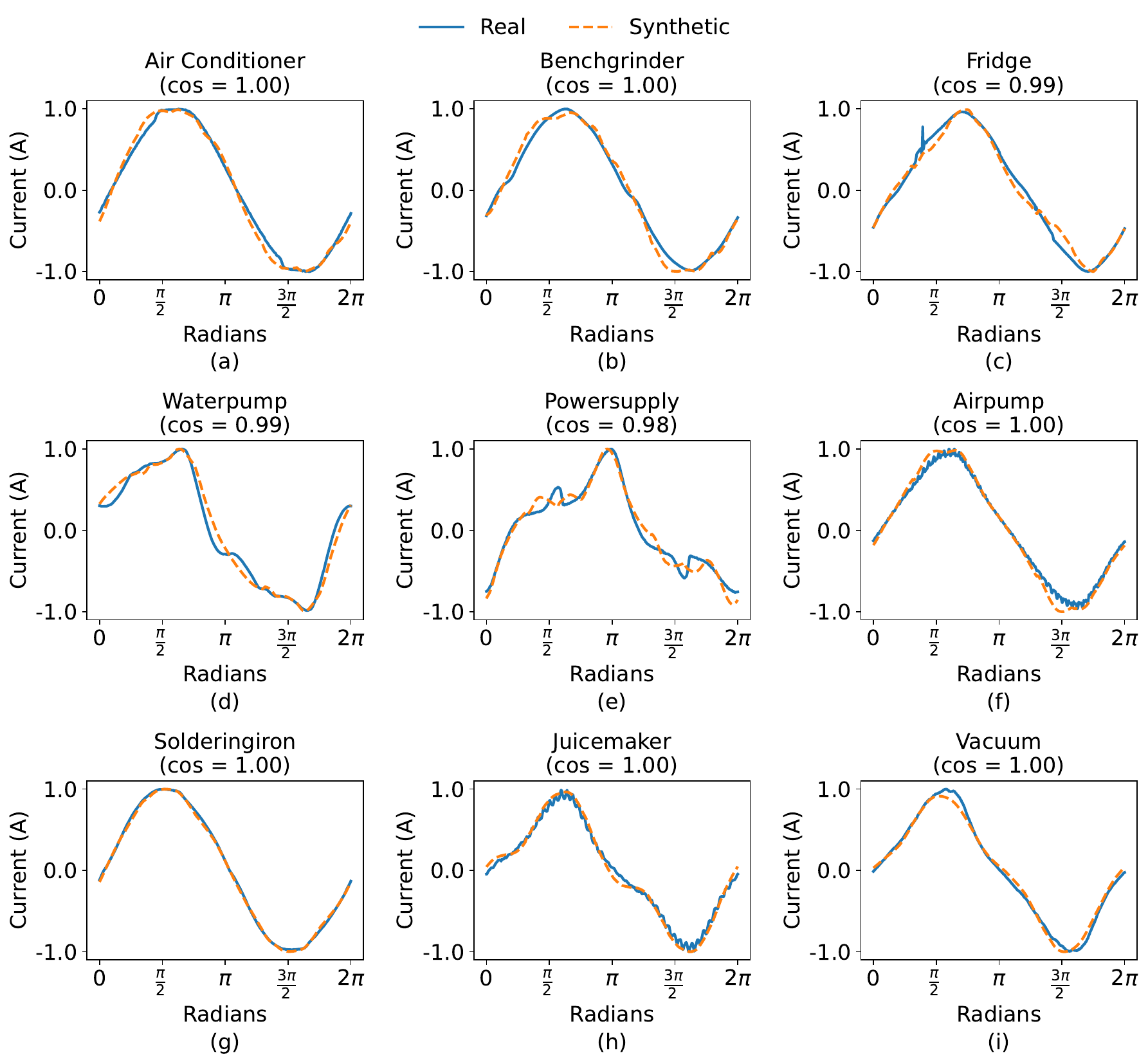}
  \caption{Comparison of arbitrary cycles from synthetic and real waveforms across nine appliance classes. High cosine similarity values ($ \ge 0.98$) indicate the high fidelity of HiFAKES, with synthetic waveforms that are visually and statistically indistinguishable from real ones.}
  \label{fig:real-syn_cos}
\end{figure}

We further evaluate the model’s fidelity by comparing waveforms of individual cycles. We randomly select nine real appliance classes, extract one arbitrary cycle per class, and match it with the most similar synthetic cycle using cosine similarity. Fig.~\ref{fig:real-syn_cos} shows these real-synthetic pairs, where cosine similarities often exceed $0.99$, and the synthetic waveforms are visually indistinguishable from the real ones.

\begin{figure}[ht!]
  \centering
  \includegraphics[width=1.\textwidth]{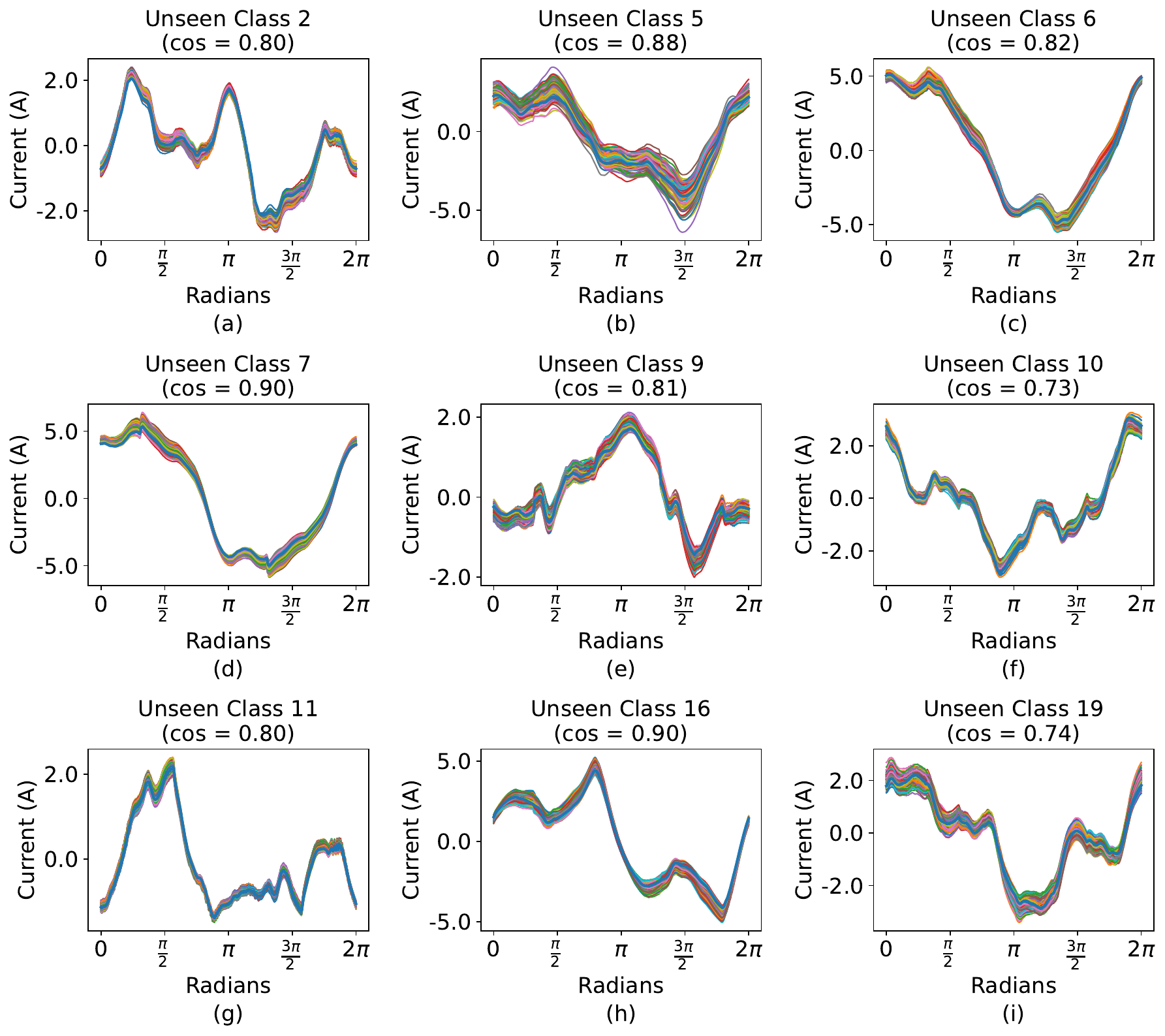}
  \caption{Bundles of synthetic cycles for nine unseen appliance classes.}
  \label{fig:unseen}
\end{figure}

\begin{figure}[ht!]
  \centering
  \includegraphics[width=1.\textwidth]{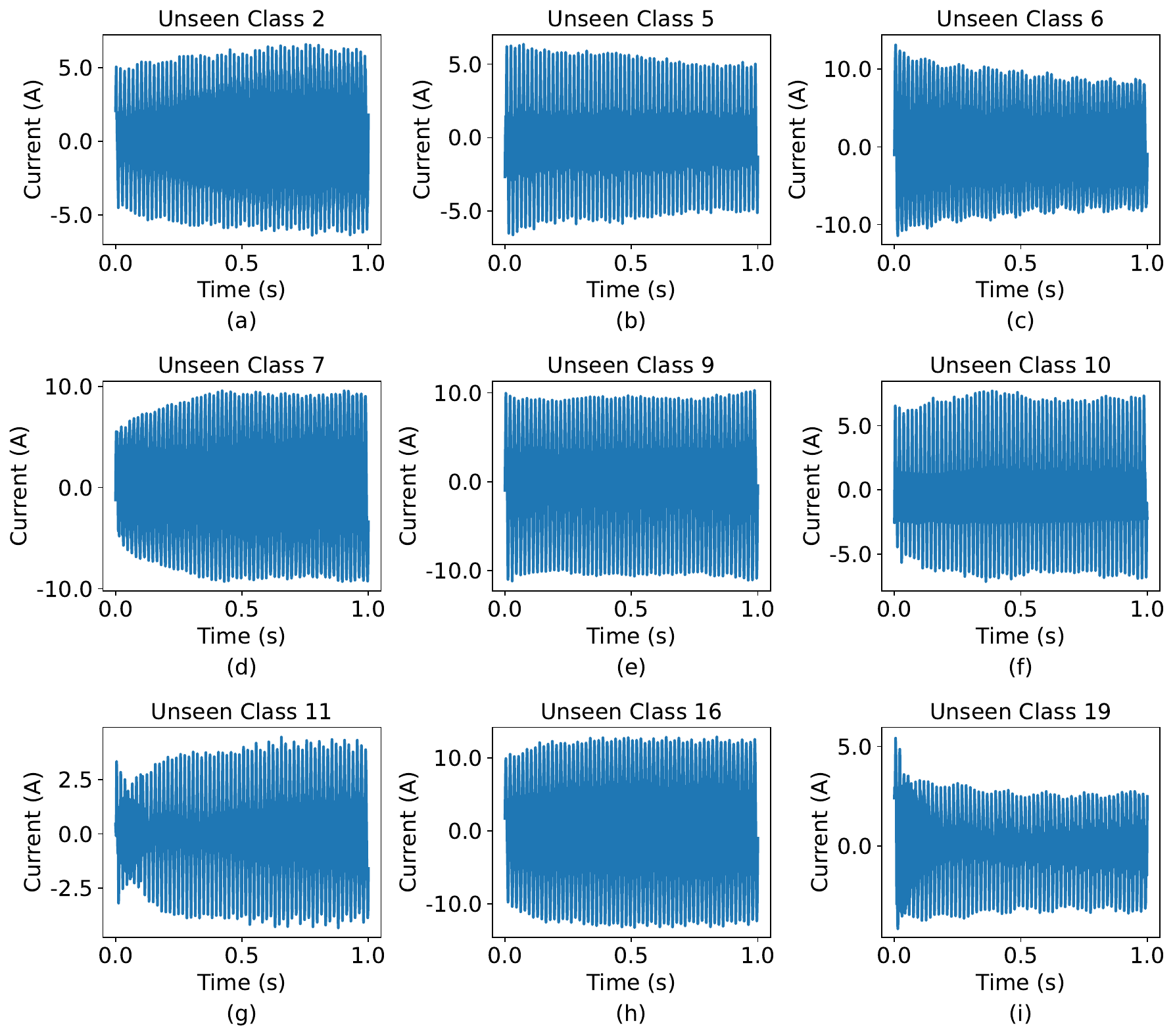}
  \caption{One-second synthetic transients generated by HiFAKES for randomly chosen classes of unseen appliances (a-i) demonstrating realistic switching and temporal dynamics.}
  \label{fig:unseen-transients}
\end{figure}

To test HiFAKES' ability to generalize to unseen appliances, we randomly sample synthetic cycles with cosine similarity to real data in the range $[0.70, 0.90]$, selecting those that lie outside high-density regions of the real distribution. For visualization purpose, we ensure that each selected synthetic cycle corresponds to a distinct appliance or its state. Figure~\ref{fig:unseen} shows bundles of cycles corresponding to each unseen appliance with little variations in amplitude and shape as typically happens right after switching event. These unseen cycles demonstrate waveforms that visually differ from real patterns but still resemble common features with the real ones: dominance of the first harmonic, phase shift in the range $[-\pi/2,\pi/2]$, exponential decay of harmonics amplitude with their order growing \cite{piasg}, see Fig.~\ref{fig:unseen} and Fig.~\ref{fig:unseen-transients} for reference.

Finally, to examine temporal consistency beyond isolated cycles, we present complete one-second transients generated by HiFAKES in Fig.~\ref{fig:unseen-transients}. These visualizations illustrate how HiFAKES simulates switching events and temporal dynamics observed in real current signatures.

\subsection{3D-metric test}

\begin{figure}[ht!]
  \centering
  \includegraphics[width=0.75\textwidth]{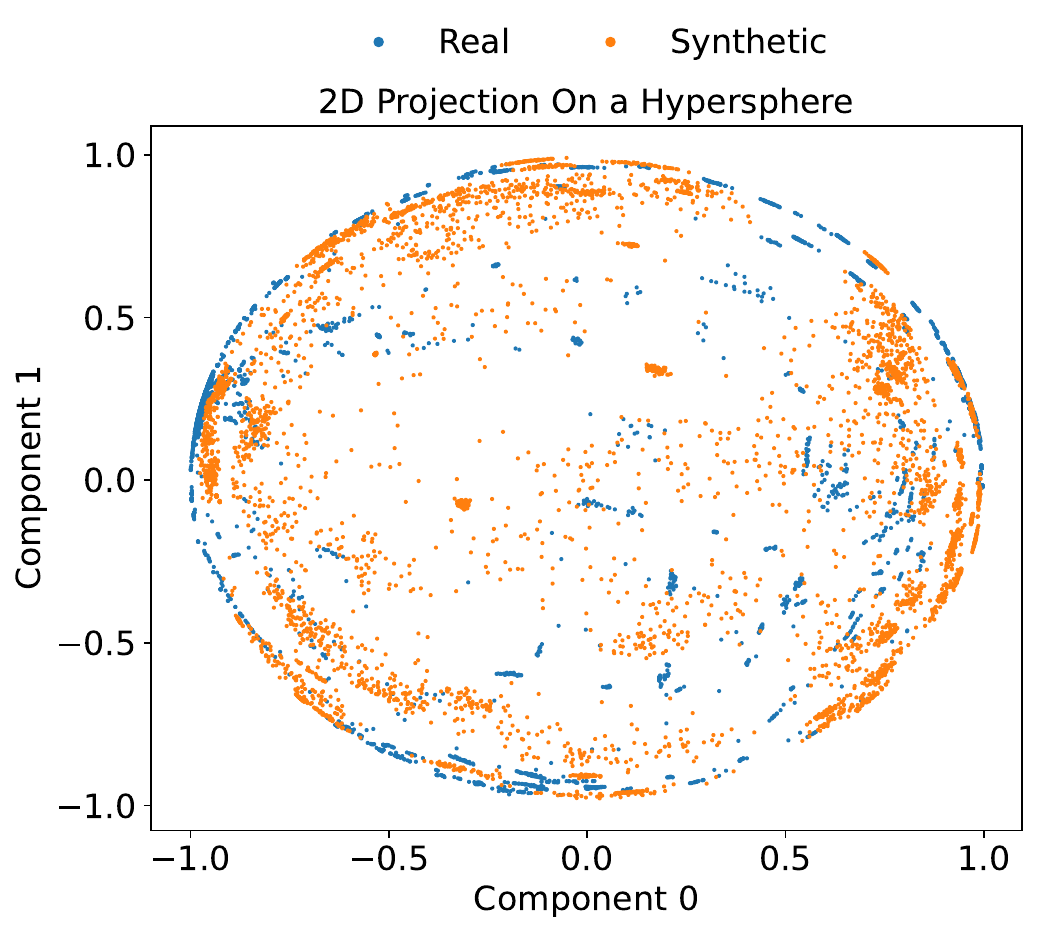}
  \caption{2D projection of real and synthetic signatures onto a hypersphere. The overlap between the two sets indicates that the data generated by HiFAKES has similar geometric structure to real appliance signatures.}
  \label{fig:hypersphere}
\end{figure}

To numerically evaluate the quality of generated data, we adopt the 3D-metric proposed in \cite{alaa2022faithful}, which consists of \(\alpha\)-precision (\( P_{\alpha} \)), \(\beta\)-recall (\( R_{\beta} \)), and authenticity. The metric generalizes earlier notions of precision and recall \cite{flach2015precision} by introducing density-based support regions in the data space. \( P_{\alpha} \) measures the proportion of generated samples that lie within the high-density region of the real distribution (the \(\alpha\)-support), while \( R_{\beta} \) quantifies the proportion of real samples covered by the \(\beta\)-support of the generative distribution. Authenticity assesses whether synthetic samples are indistinguishable from real samples by comparing nearest-neighbor distances. To compute the $\alpha$-precision $P_{\alpha}$ and $\beta$-recall $R_{\beta}$, one should first project the real dataset $X_r$ and the synthetic dataset $X_g$ onto minimum-volume hyper-spheres $\mathcal{S}_r$ and $\mathcal{S}_g$, with centers $o_r$ and $o_g$, respectively. Here, we follow the approach presented in \cite{alaa2022faithful} to project  $X_r$ and $X_g$ onto hyper-spheres (see Fig.~\ref{fig:hypersphere}).

By taking $\alpha$ and $\beta$ quantiles of the radii of the mentioned spheres, the estimates of $\alpha$- and $\beta$-supports can be introduced. Thus, the $\alpha$-support is a concentric Euclidean ball $\mathcal{S}^{\alpha}_r = B(o_r, r_{\alpha})$ with radius $r_{\alpha} = Q_{\alpha}(\|\tilde{x}_r - o_r\| : \tilde{x}_r \in \widetilde{X}_r)$, where $Q_{\alpha}$ is a function that computes the $\alpha$ quantile, and $\widetilde{X}_r = \Phi(X_r)$ is the projection of $X_r$. The $\beta$-support is defined in the same fashion i.e., $\widetilde{X}_g=\Phi(X_g)$. It is assumed that data points outside of the $\alpha$- or $\beta$-support are outliers. The $\alpha$-precision is defined as a probability that a synthetic sample resides in the $\alpha$-support:
\begin{equation}
    P_{\alpha}=\frac{\sum\mathbf{1}(d(\widetilde{x}_g, o_r)\leq r_{\alpha})}{|\widetilde{X}_g|}~~~\text{for}\,\alpha\in [0;1]
    \label{eq:P_alpha}
\end{equation}
where $\mathbf{1}$ is the function that indicates 1 if the condition in the parentheses is true, and 0 otherwise, $d$ is the Euclidean distance, and the summation is done over all $\tilde{x}_g\in\widetilde{X}_g$, and $P_{\alpha}\in [0;1]$.

The $\beta$-recall is defined as a probability that each real sample is locally covered by the nearest synthetic sample from the $\beta$-support:
\begin{equation}
    R_{\beta}=\frac{\sum\mathbf{1}(\tilde{x}^*_{g,\beta}\in B(\widetilde{x}_r, d(\tilde{x}_r, \tilde{x}^*_r)))}{|\widetilde{X_r}|}~~~\text{for}\,\beta\in [0;1]
    \label{eq:R_beta}
\end{equation}
where \( \tilde{x}^*_{g,\beta} \) is a synthetic sample belonging to the \(\beta\)-support, which is nearest to a real sample \( \tilde{x}_r \). \( B \) is a Euclidean ball centered at the real sample \( \tilde{x}_r \) with a radius equal to the distance to its \( k \)-th nearest neighbor \( \tilde{x}^*_r \). The summation is done over all \( \tilde{x}_r \in \widetilde{X}_r \), and \( R_{\beta} \in [0;1] \).

The authenticity is defined as:
\begin{equation}
    A=\frac{\sum\textbf{1}(d(\tilde{x}_r, \tilde{x}^*_r)<d(\tilde{x}_r,\tilde{x}^*_g))}{|\widetilde{X}_g|}
    \label{eq:A}
\end{equation}
where $\tilde{x}^*_g$ is the nearest synthetic sample to a real sample $\tilde{x}_r$, and $\tilde{x}^*_r$ is the nearest real sample to $\tilde{x}_r$, and summation is done over all $\tilde{x}_r\in\widetilde{X}_r$, and $A\in [0;1]$.

\begin{figure}[ht!]
  \centering
  \includegraphics[width=1\textwidth]{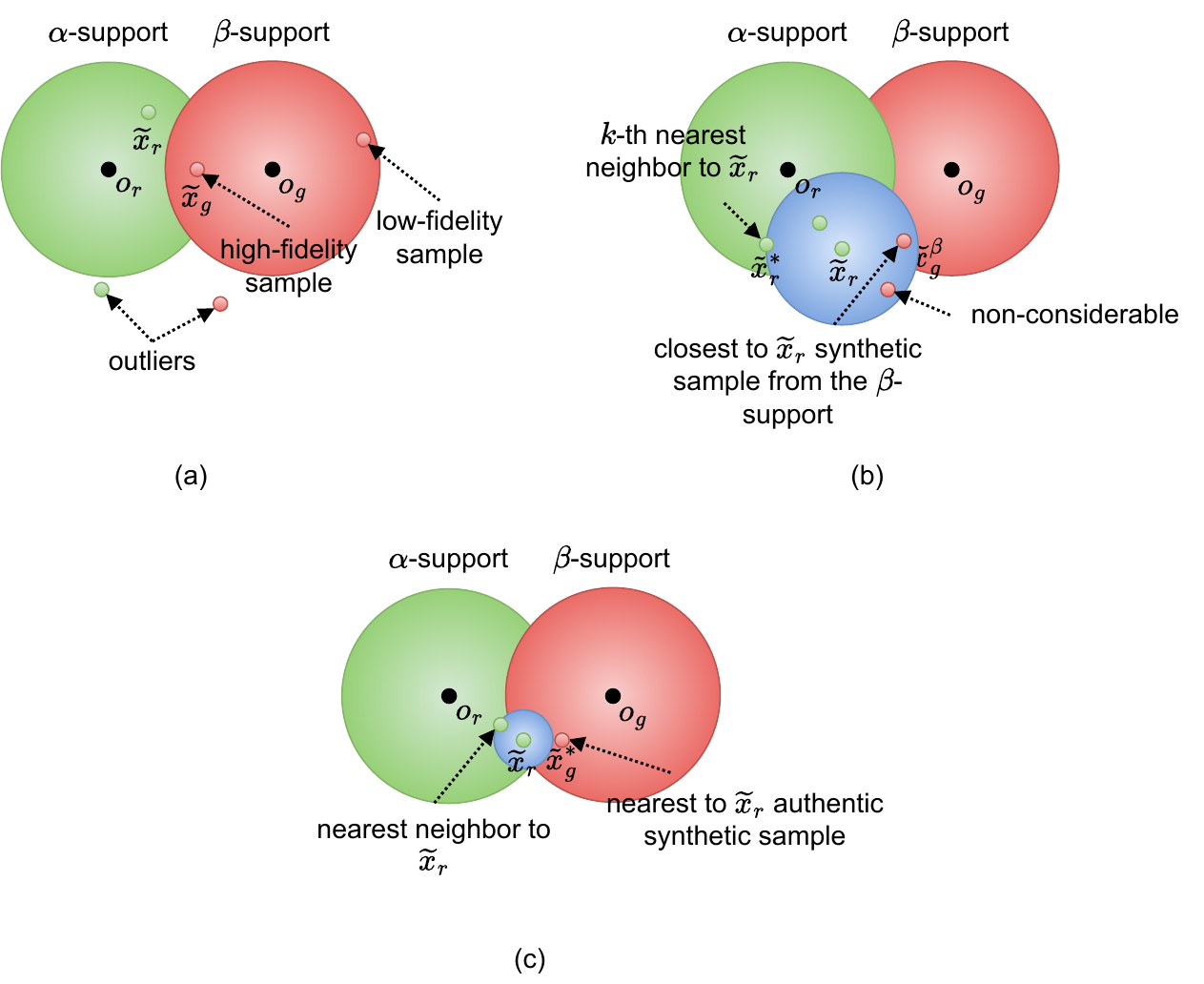}
  \caption{Graphical representation of the key variables used for the computing of $\alpha$-precision (a), $\beta$-recall (b), and authenticity (c).}
  \label{fig:balls}
\end{figure}

It is recommended to use integrated metrics $\alpha$-precision and $\beta$-recall to assess the performance of a generative model in a single number \cite{flach2015precision} :
\begin{equation}
\begin{gathered}
    IP_{\alpha}=1-2\Delta P_{\alpha},
    IR_{\beta}=1-2\Delta R_{\beta},
    \\
    \Delta P_{\alpha}=\int_0^1|P_{\alpha}-\alpha|d\alpha,\Delta R_{\beta}=\int_0^1|R_{\beta}-\beta|d\beta,
\end{gathered}
\end{equation}
where $IP_{\alpha}\in[0;1]$, and $IR_{\beta}\in [0;1]$. The ideal situation occurs when the real and generative distributions are equal, corresponding to $IP_{\alpha}=IR_{\beta}= 1$. For a better understanding of the logic behind equations \ref{eq:P_alpha}, \ref{eq:R_beta} and \ref{eq:A}, we provide illustrations for each formula in Fig.~\ref{fig:balls}. For further details, one can refer to the original paper \cite{alaa2022faithful}.

\begin{figure}[ht!]
  \centering
  \includegraphics[width=0.66\textwidth]{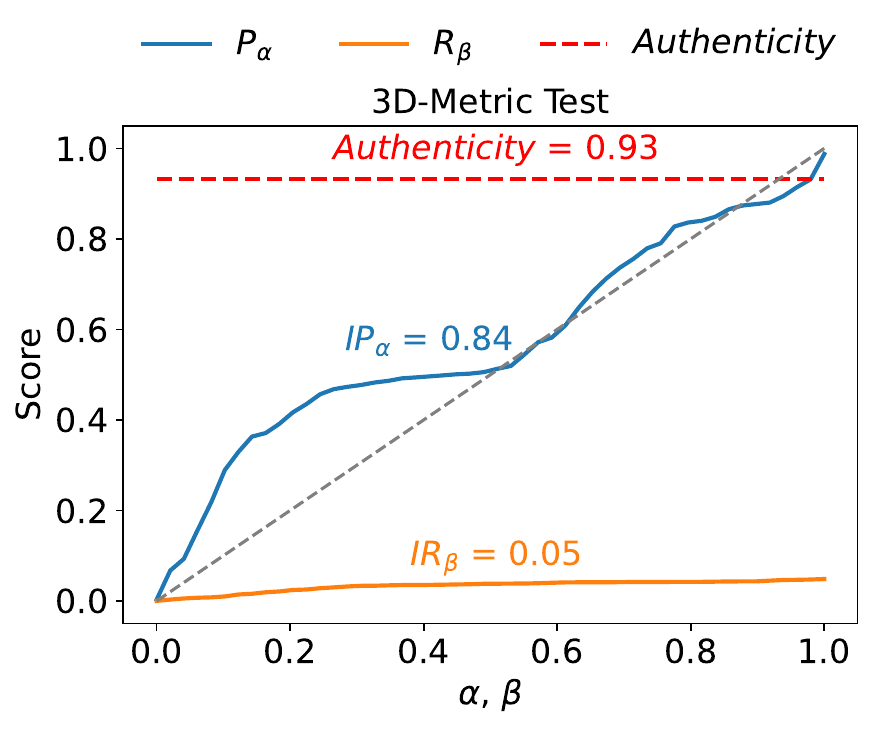}
  \caption{3D-metric test for HiFAKES: integrated $\alpha$-precision, integrated $\beta$-recall and authenticity.}
\label{fig:3d-metric}
\end{figure}

As shown in Fig.~\ref{fig:3d-metric}, HiFAKES achieves a high authenticity score of $93\%$, an integrated $\alpha$-precision ($IP_{\alpha}$) of $84\%$, and an integrated $\beta$-recall ($IR_{\beta}$) of $5\%$. 
These results indicate that the generated submetered signatures are highly realistic and well-aligned with the dominant modes of the true distribution. Although the recall appears modest, it is consistent with the range reported for generative models in other application domains, as evaluated in \cite{alaa2022faithful}, where $IR_{\beta}$ values ranged from $0.3\%$ to $17\%$ on the AmsterdamUMCdb dataset. As emphasized in the original study, a low $IR_{\beta}$ does not necessarily indicate poor model quality, but rather reflects the inherent difficulty of fully covering complex, high-entropy data distributions.

\section{Case Study: Controlled Evaluation of Generalization in NILM}
\label{sec:case}

NILM must perform reliably under diverse and often challenging real-world conditions. Both the choice of features and the learning model play a critical role in how well a system generalizes to unseen appliances, and signatures complexities. Each model together with features selected can behave differently depending on factors such as how distinct appliance signatures are, how many devices operate simultaneously, and whether the model has seen examples from a particular brand.

This case study demonstrates how HiFAKES enables benchmarking of NILM models subject to each of these factors independently: feature separability, simultaneous appliance operation, and brand variability. For each experiment, we vary a single condition while holding others fixed, allowing us to isolate its impact on disaggregation accuracy.

To benchmark performance, we use three standard machine learning models chosen for their popularity as \textit{baseline} models in a prior NILM work~\cite{features}, and interpretability: K-Nearest Neighbors (KNN), decision tree, and XGBoost. We use each model as a multi-output regressor to solve the power estimation problem, i.e., to identify fractions of power consumed by each active appliance. 

Disaggregation performance is measured using the coefficient of determination ($R^2$), which quantifies how accurately the predicted appliance-level energy fractions match the true values. An $R^2$ of 1.0 indicates perfect prediction; values closer to zero or negative imply weak or misleading outputs.

This study can be extended to the models beyond the selected ones and therefore guide the researchers to identify which models are most effective under which conditions something that is difficult to do with real-world data alone.

\subsection{Selected Features}

The feature set used in our analysis directly follows prior comprehensive evaluations in NILM literature~\cite{features}, specifically chosen to capture diverse and discriminative characteristics of appliance signatures. The following features were computed from each generated aggregated current signature $\mathbf{X}_a^{(i)}$ and the reference grid voltage signature $\mathbf{v}$:

\begin{enumerate}
  \item \textbf{Form Factor} (FF) is a measure of waveform distortion and it is defined as the ratio of the root mean square (RMS) to the mean of the absolute current:
  \begin{equation}
      \mathrm{FF} = \frac{\mathrm{rms}(\mathbf{X}_a^{(i)})}{\mathrm{mean}(|\mathbf{X}_a^{(i)}|)}   
  \end{equation}

  It captures the harmonic content and helps in distinguishing nonlinear devices like chargers from linear devices like heater.

  \item \textbf{Temporal Centroid} (TC) reflects the temporal energy distribution of the current waveform over $P$ periods:
  
\begin{equation}
    \mathrm{TC} = \frac{\sum_{p=1}^P \mathbf{\widehat{X}}_a^{(i,p)} \cdot p}{\sum_{p=1}^P \mathbf{\widehat{X}}_a^{(i,p)}}
\end{equation}

  where $\mathbf{\widehat{X}}_a$ is the tensor of dimensions $A\times P\times \lfloor T/f_0\rfloor$ and $f_0=60$ Hz is the mains frequency of PLAID and WHITED datasets. It is effective for capturing transient-rich behavior typical for motor-driven or variable-power appliances.

  \item \textbf{Admittance Over Time} (AOT) captures the time-varying admittance of the appliance, calculated as:
\begin{equation}
   \mathrm{AOT}^{(p)} = \frac{\mathbf{\widehat{X}}_a^{(i,p)}}{\mathbf{v}}, \quad p = 1, \dots, P
\end{equation}

  It highlights the reactive components in load behavior.

  \item \textbf{Wavelet Energy} (WE) computes the energy in wavelet decomposition bands:
\begin{equation}
    \mathrm{WE}^{(j)} = \sum_{n} |\mathbf{D}_n^{(j)}|^2
\end{equation}

  where $\mathbf{D}^{(j)}$ are wavelet coefficients at level $j$ of the discrete wavelet transform. This feature captures both spectral and transient characteristics across multiple scales.

  \item \textbf{VI Trajectory} (VIT) is a sampled 2D path of the normalized voltage and current waveforms:
  \begin{equation}
      \mathrm{VIT} = \left\{ \left( \frac{\mathbf{v}^{(t)}}{\max(|\mathbf{v}^{(t)}|)}, \frac{\mathbf{X}_a^{(i,t)}}{\max(|\mathbf{X}_a^{(i,t)}|)} \right) \right\}_{t=1}^{T}
  \end{equation}

It visualizes nonlinear load behaviors and provides discriminative signatures for devices based on their voltage-current interaction.

\item \textbf{Phase Shift} measures the phase angle \(\theta\) between voltage and current:
\begin{equation}
  \theta = \cos^{-1}\left(\frac{W}{S}\right)
\end{equation}

The active power \(W\) is computed as the mean of the element-wise product ($\odot$) between the aggregated current \(\mathbf{X}_a^{(i)}\) and the voltage signal \(\mathbf{v}\):
\begin{equation}
  W = \text{mean}\left(\mathbf{X}_a^{(i)} \odot \mathbf{v}\right)
\end{equation}

The apparent power \(S\) is calculated as follows:
\begin{equation}
  S = \text{rms}\left(\mathbf{X}_a^{(i)}\right) \odot \text{rms}\left(\mathbf{v}\right)
\end{equation}

The phase shift is a useful scalar feature that distinguishes between resistive, inductive, and capacitive appliances.

\end{enumerate}

All the above features collectively ensure a comprehensive characterization of temporal, spectral, and nonlinear properties of appliance signatures for robust appliance identification and power estimation.

\subsection{Experiment 1: Class Separability}

In the framework of HiFAKES, we introduce class separability ($\varepsilon_{\text{sep}}$), which refers to the degree of distinction among different appliance classes in the synthetic PCA space. High class separability indicates that generated appliances have distinct signatures, which might result in accurate power estimation or load identification. This property is especially critical in NILM where many devices may exhibit similar temporal or spectral characteristics, e.g., water kettle and boiler, or laptop charger and LED lighting.

Evaluating class separability provides an early indication of the discriminative power of both the selected features and the NILM model. In this experiment, we varied the class separability, defined as the mean distance between appliance classes in the synthetic PCA space obtained at the second stage of HiFAKES framework (see Fig.~\ref{fig:framework}), from completely overlapping ($\varepsilon_{\text{sep}}=0.0$) to well-separated ($\varepsilon_{\text{sep}}=2.0$).

\begin{figure}[ht!]
    \centering
    \includegraphics[width=0.66\textwidth]{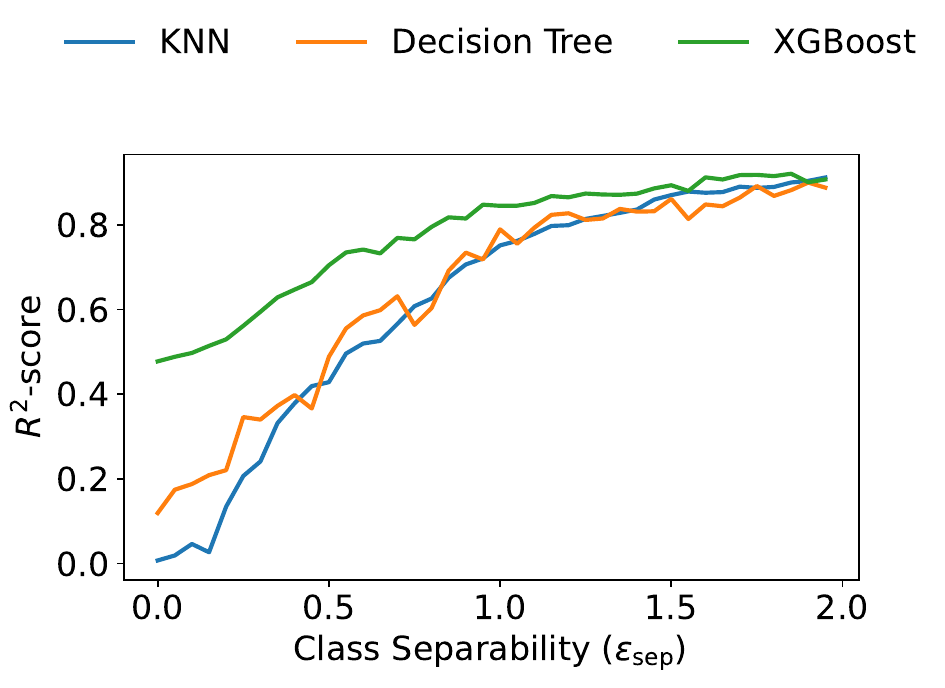}
    \caption{Impact of feature separability ($\varepsilon_{\text{sep}}$) on NILM performance. As $\varepsilon_{\text{sep}}$ increases from 0.0 to 2.0, the $R^{2}$ scores improve across all models, with XGBoost outperforming the others.}
    \label{fig:sep}
\end{figure}

\begin{figure}[ht!]
    \centering
    \includegraphics[width=1.0\textwidth]{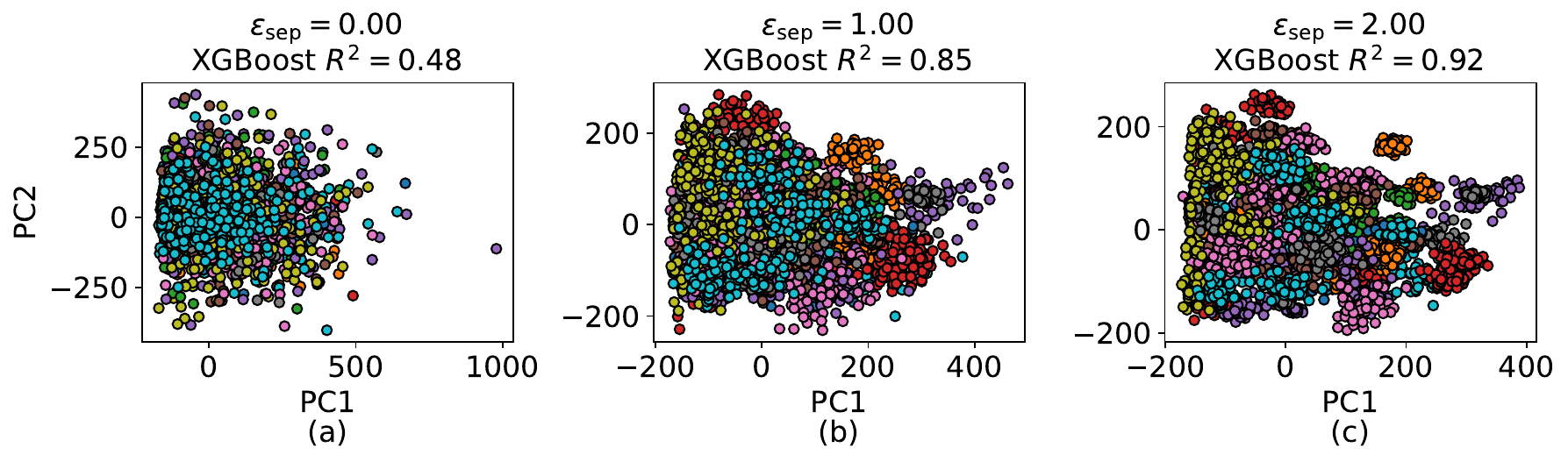}

    \caption{Impact of class separability ($\varepsilon_{\text{sep}}$) on the disaggregation performance of the best performing model, XGBoost. As $\varepsilon_{\text{sep}}$ increases, clusters become more distinct, and the $R^{2}$ score of XGBoost improves from 0.48 to 0.92, indicating better energy disaggregation.}
    \label{fig:boundaries}
\end{figure}

As shown in Fig.~\ref{fig:sep}, increasing separability directly improves power estimation accuracy. At minimal separability (0.0), all selected classifiers yield poor performance. With higher separability (2.0), XGBoost achieves an $R^2$ of 0.81, outperforming decision tree (0.73) and KNN (0.66). The results also highlight that certain NILM models (as XGBoost in our experiment), are more robust to the situations when class boundaries are less distinct. A visual example of the best performing model (XGBoost) is presented in Fig.~\ref{fig:boundaries}.

This experiment demonstrates the utility of synthetic benchmarking for model selection. By varying separability, it is possible to identify which NILM models potentially will lead to better generalization on real-world data. To repeat the same experiment on real-world data, a large amount of novel signatures beyond the existing datasets must be collected. Synthetic data thus serves as a controlled and scalable alternative. 

\subsection{Experiment 2: Simultaneously operating appliances}

In real households, appliances rarely operate in isolation. At any given moment, several devices such as the refrigerator, Wi-Fi router, and water boiler are typically running simultaneously that create an aggregate current signal composed of multiple steady-state loads. A statistical analysis in~\cite{kamyshev2025coldconcurrentloadsdisaggregator} shows that 3 to 5 appliances are active simultaneously most of the time. This constant background load presents a challenge for NILM systems, especially when relying on event-based detection, which assumes clean on/off transitions.

\begin{figure}[htbp]
    \centering
    \includegraphics[width=0.9\textwidth]{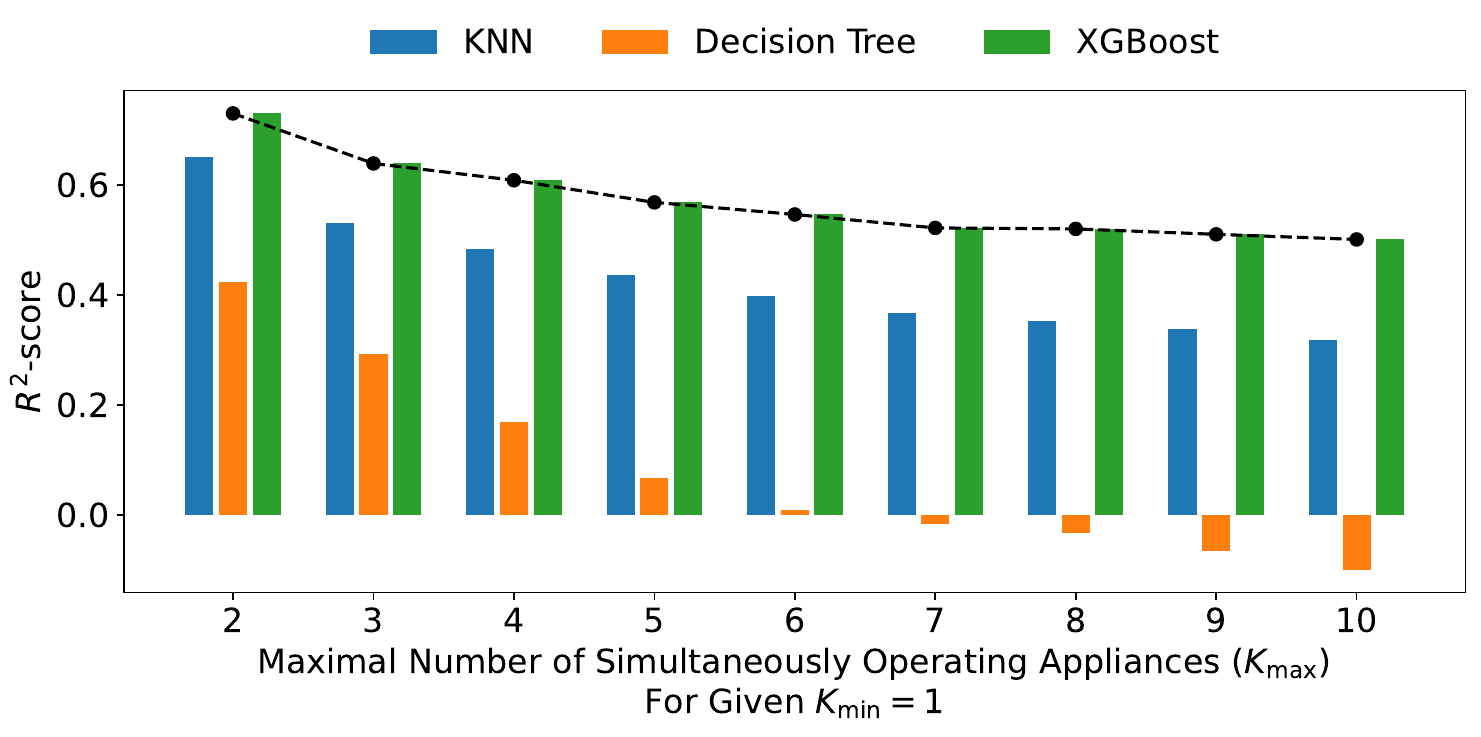}
    \caption{Effect of appliance concurrency on disaggregation accuracy. The higher the number of appliances which are running at the same time, the lower the accuracy of all models.}
    \label{fig:simult}
\end{figure}

This experiment evaluates how NILM models perform when multiple appliances operate at the same time. This will guide in choosing NILM models that are most robust to the number of active appliances. Unlike Experiment 1, here we generate \textit{aggregate} synthetic current signatures for the whole range of 2 to 10 randomly selected active appliances. Figure~\ref{fig:simult} shows that disaggregation accuracy declines as more appliances are on, but not uniformly across models. Similarly, XGBoost remains the most stable, with an $R^2$ of 0.61 at five active appliances.  In contrast, the decision tree model performs the worst, with scores dropping below zero when more than 6 appliances are active.

This particular task highlights a key aspect of real-world NILM: appliances do not operate in isolation. Synthetic evaluation under controlled multi-load conditions helps reveal which NILM models can scale to realistic usage without requiring precise event locations.

\subsection{Experiment 3: Brand generalization}

Appliances of the same class often differ significantly across brands mainly due to variations in internal circuitry, control logic, and power electronics. A model trained on one brand may not necessarily generalize well to another one. This experiment evaluates how well NILM models trained on one subset of brands perform when tested on the unseen subset of brands from the same appliance class. For consistency, we sample subsets of brands for each class of appliances.

To conduct this experiment, we split each appliance class into disjoint sets of brands for training and testing, where the ratio $\tau$ denotes the share of brands included in the training set, and $1 - \tau$ corresponds to those kept out for testing. All samples in the test set come from brands never seen during training. This setup reflects a realistic deployment scenario, where the NILM model encounters new devices not represented in its training data. HiFAKES enables this evaluation by assigning multiple synthetic brands per class, each generated using KMeans clustering in the synthetic PCA space prior to converting synthetic signature into the time-domain. 

\begin{figure}[htbp!]
    \centering
    \includegraphics[width=0.66\textwidth]{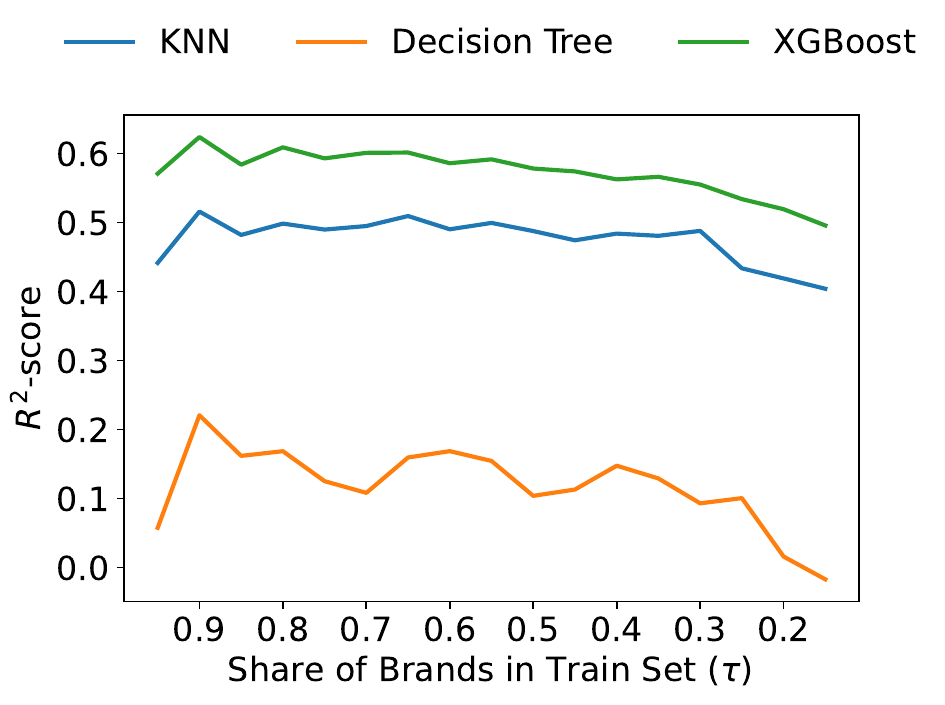}
    \caption{Impact of brand diversity on model generalization. Here, $\tau$ denotes the fraction of appliance brands used for training. As $\tau$ decreases, the R² score drops for all models, indicating reduced generalization to unseen brands. XGBoost remains the most robust across all values of $\tau$.}
    \label{fig:brand}
\end{figure}

Figure~\ref{fig:brand} shows a consistent drop in performance as the model is trained on fewer appliance brands (i.e., as $\tau$ decreases). XGBoost again performs best, achieving an average $R^2$ of around 0.58, while KNN remains relatively stable around 0.47, and Decision Tree drops below 0.05. These results confirm that brand variation is a non-trivial challenge and must be explicitly addressed to ensure NILM model robustness.

Thus, this test provides practical guidance on how much brand diversity is needed for training a particular model. As shown in Figure~\ref{fig:brand}, the drop in performance between 90\% and 50\% of training brands is relatively small, suggesting that models can still generalize well even with limited brand coverage. For example, XGBoost maintains strong performance when trained on just 90\% of brands, indicating that it may require less data to handle unseen devices effectively. However, the optimal training share may vary across models, and this type of evaluation should be repeated when selecting or designing new NILM models.

\section{Conclusion}
\label{sec:dis}

NILM requires a plug-and-play data generator capable of producing synthetic datasets that are realistic, diverse, and fully customizable. While previous methods can increase training diversity, they are not designed for: (i) generating fully unseen testing scenarios for debugging and finding a robust NILM model to the unseen real-world scenarios; (ii) enabling rapid prototyping; (iii) providing full control over submetered and aggregate signatures. Our approach fills this gap, and serves a similar role for high-frequency NILM as $\texttt{make\_classification}$ does in the Python library named \textit{scikit-learn}. 

In this paper, we introduced HiFAKES, a high-frequency synthetic data generator designed to address a critical gap in NILM: the lack of diverse, labeled, and scalable datasets for model diagnosis and generalization evaluation. Through extensive experiments and theoretical analysis, we demonstrated that HiFAKES enables controlled, reproducible testing of NILM models across key real-world challenges—appliance similarity, simultaneous operation, and brand diversity—while requiring no additional real-world data collection.

HiFAKES is computationally efficient and lightweight, requiring only 25 seconds to train on a standard CPU. To further lower the entry barrier for researchers and practitioners, we made a pre-trained version available on GitHub. This allows immediate use of the generator without access to any real-world dataset, facilitating rapid prototyping, model debugging, and controlled benchmarking from the outset.

Our evaluation shows that HiFAKES produces high-fidelity appliance signatures that are statistically and visually similar to real-world data. The 3D-metric test confirms the authenticity and diversity of the generated samples, achieving an integrated $\alpha$-precision ($IP_{\alpha}$) of 84\% and an authenticity score of 93\%. Visual comparisons further support the realism of synthetic waveforms, capturing essential physical traits such as waveform symmetry, harmonic decay, and phase alignment.

A key strength of HiFAKES lies in its flexibility. Unlike prior generative or augmentation methods, it offers fine-grained control over generation parameters such as class separability, intra-class diversity, number of brands and modes, and concurrent appliance operation. This makes HiFAKES valuable for stress-testing NILM models in challenging and customizable real-world scenarios. 

The results obtained with our case study confirm that synthetic evaluation is a powerful tool for identifying robust NILM models and feature sets prior to deployment. For example, XGBoost consistently outperforms simpler models such as KNN and decision trees when facing overlapping appliance classes or unseen brands. These insights would be difficult to obtain using only limited or noisy real-world datasets.

One limitation of HiFAKES is that it is only intended for short-term data (one-minute window); in order to provide a more complete solution, long-term time periods, like hours or days, must be generated. In this instance, a complete load profile can also be produced artificially, which will create new possibilities for NILM as well as other fields of study such as load forecasting and demand response. Therefore, extending the duration of generated cycles will be the primary focus of future works.


\section*{Acknowledgements}
The authors acknowledge partial support by the Skoltech program: Skolkovo Institute of Science and Technology – Hamad Bin Khalifa University Joint Projects. 

\newpage
 \bibliographystyle{elsarticle-num} 
 \bibliography{cas-refs}

\begin{thebibliography}{10}
\expandafter\ifx\csname url\endcsname\relax
  \def\url#1{\texttt{#1}}\fi
\expandafter\ifx\csname urlprefix\endcsname\relax\def\urlprefix{URL }\fi
\expandafter\ifx\csname href\endcsname\relax
  \def\href#1#2{#2} \def\path#1{#1}\fi

\bibitem{strielkowski2021renewable}
W.~Strielkowski, L.~Civ{\'\i}n, E.~Tarkhanova, M.~Tvaronavi{\v{c}}ien{\.e},
  Y.~Petrenko, Renewable energy in the sustainable development of electrical
  power sector: A review, Energies 14~(24) (2021) 8240.

\bibitem{iea2}
{International Energy Agency}, Electricity 2025,
  \url{https://www.iea.org/reports/electricity-2025}, iEA, Paris. Licence: CC
  BY 4.0 (2025).

\bibitem{locmelis2020benchmarking}
K.~Locmelis, D.~Blumberga, A.~Blumberga, A.~Kubule, Benchmarking of industrial
  energy efficiency. outcomes of an energy audit policy program, Energies
  13~(9) (2020) 2210.

\bibitem{barbar2023impact}
M.~Barbar, D.~S. Mallapragada, R.~J. Stoner, Impact of demand growth on
  decarbonizing india's electricity sector and the role for energy storage,
  Energy and Climate Change 4 (2023) 100098.

\bibitem{umbark2020energy}
M.~A. Umbark, S.~K. Alghoul, E.~I. Dekam, Energy consumption in residential
  buildings: Comparison between three different building styles, Sustainable
  Development Research 2~(1) (2020) p1--p1.

\bibitem{IEA_2024}
IEA, \href{https://www.iea.org/reports/world-energy-outlook-2024}{World energy
  outlook 2024 – analysis} (2024).
\newline\urlprefix\url{https://www.iea.org/reports/world-energy-outlook-2024}

\bibitem{savaresi2016paris}
A.~Savaresi, The paris agreement: a new beginning?, Journal of Energy \&
  Natural Resources Law 34~(1) (2016) 16--26.

\bibitem{ruano2019nilm}
A.~Ruano, A.~Hernandez, J.~Ure{\~n}a, M.~Ruano, J.~Garcia, Nilm techniques for
  intelligent home energy management and ambient assisted living: A review,
  Energies 12~(11) (2019) 2203.

\bibitem{YOUSSEF2023127793}
H.~Youssef, S.~Kamel, M.~H. Hassan, L.~Nasrat, Optimizing energy consumption
  patterns of smart home using a developed elite evolutionary strategy
  artificial ecosystem optimization algorithm, Energy 278 (2023) 127793.

\bibitem{BECKEL2014397}
C.~Beckel, L.~Sadamori, T.~Staake, S.~Santini, Revealing household
  characteristics from smart meter data, Energy 78 (2014) 397--410.
\newblock \href {https://doi.org/https://doi.org/10.1016/j.energy.2014.10.025}
  {\path{doi:https://doi.org/10.1016/j.energy.2014.10.025}}.

\bibitem{pereira2018performance}
L.~Pereira, N.~Nunes, Performance evaluation in non-intrusive load monitoring:
  datasets, metrics, and tools—a review, Wiley Interdisciplinary Reviews:
  data mining and knowledge discovery 8~(6) (2018) e1265.

\bibitem{Hart1992NonintrusiveAL}
G.~W. Hart, Nonintrusive appliance load monitoring, Proceedings of the IEEE
  80~(12) (1992) 1870--1891.

\bibitem{schirmer2022non}
P.~A. Schirmer, I.~Mporas, Non-intrusive load monitoring: A review, IEEE
  Transactions on Smart Grid 14~(1) (2022) 769--784.

\bibitem{Event_detection}
K.~D. Anderson, M.~E. Bergés, A.~Ocneanu, D.~Benitez, J.~M. Moura, Event
  detection for non intrusive load monitoring, in: IECON 2012 - 38th Annual
  Conference on IEEE Industrial Electronics Society, 2012, pp. 3312--3317.
\newblock \href {https://doi.org/10.1109/IECON.2012.6389367}
  {\path{doi:10.1109/IECON.2012.6389367}}.

\bibitem{CARRIEARMEL2013213}
K.~C. Armel, A.~Gupta, G.~Shrimali, A.~Albert, Is disaggregation the holy grail
  of energy efficiency? the case of electricity, Energy policy 52 (2013)
  213--234.

\bibitem{azad2023non}
M.~I. Azad, R.~Rajabi, A.~Estebsari, Non-intrusive load monitoring (nilm) using
  deep neural networks: A review, in: 2023 IEEE International Conference on
  Environment and Electrical Engineering and 2023 IEEE Industrial and
  Commercial Power Systems Europe (EEEIC/I\&CPS Europe), IEEE, 2023, pp. 1--6.

\bibitem{kelly2015neural}
J.~Kelly, W.~Knottenbelt, Neural nilm: Deep neural networks applied to energy
  disaggregation, in: Proceedings of the 2nd ACM international conference on
  embedded systems for energy-efficient built environments, 2015, pp. 55--64.

\bibitem{shin2019subtask}
C.~Shin, S.~Joo, J.~Yim, H.~Lee, T.~Moon, W.~Rhee, Subtask gated networks for
  non-intrusive load monitoring, in: Proceedings of the AAAI conference on
  artificial intelligence, Vol.~33, 2019, pp. 1150--1157.

\bibitem{chen2019scale}
K.~Chen, Y.~Zhang, Q.~Wang, J.~Hu, H.~Fan, J.~He, Scale-and context-aware
  convolutional non-intrusive load monitoring, IEEE Transactions on Power
  Systems 35~(3) (2019) 2362--2373.

\bibitem{chen2018convolutional}
K.~Chen, Q.~Wang, Z.~He, K.~Chen, J.~Hu, J.~He, Convolutional sequence to
  sequence non-intrusive load monitoring, the Journal of Engineering 2018~(17)
  (2018) 1860--1864.

\bibitem{de2021recurrent}
L.~de~Diego-Ot{\'o}n, D.~Fuentes-Jimenez, {\'A}.~Hern{\'a}ndez, R.~Nieto,
  Recurrent lstm architecture for appliance identification in non-intrusive
  load monitoring, in: 2021 IEEE International Instrumentation and Measurement
  Technology Conference (I2MTC), IEEE, 2021, pp. 1--6.

\bibitem{hwang2022nonintrusive}
H.~Hwang, S.~Kang, Nonintrusive load monitoring using an lstm with feedback
  structure, IEEE Transactions on Instrumentation and Measurement 71 (2022)
  1--11.

\bibitem{kamyshev2025coldconcurrentloadsdisaggregator}
I.~Kamyshev, S.~M. Hoosh, D.~Kriukov, E.~Gryazina, H.~Ouerdane,
  \href{https://arxiv.org/abs/2106.02352}{Cold: Concurrent loads disaggregator
  for non-intrusive load monitoring} (2025).
\newblock \href {http://arxiv.org/abs/2106.02352} {\path{arXiv:2106.02352}}.
\newline\urlprefix\url{https://arxiv.org/abs/2106.02352}

\bibitem{varanasi2024stnilm}
L.~S. Varanasi, S.~P.~K. Karri, Stnilm: Switch transformer based non-intrusive
  load monitoring for short and long duration appliances, Sustainable Energy,
  Grids and Networks 37 (2024) 101246.

\bibitem{d2019transfer}
M.~D’Incecco, S.~Squartini, M.~Zhong, Transfer learning for non-intrusive
  load monitoring, IEEE Transactions on Smart Grid 11~(2) (2019) 1419--1429.

\bibitem{kaselimi2022towards}
M.~Kaselimi, E.~Protopapadakis, A.~Voulodimos, N.~Doulamis, A.~Doulamis,
  Towards trustworthy energy disaggregation: A review of challenges, methods,
  and perspectives for non-intrusive load monitoring, Sensors 22~(15) (2022)
  5872.

\bibitem{ahmed2020edge}
S.~Ahmed, M.~Bons, Edge computed nilm: a phone-based implementation using
  mobilenet compressed by tensorflow lite, in: Proceedings of the 5th
  international workshop on non-intrusive load monitoring, 2020, pp. 44--48.

\bibitem{tabanelli2021trimming}
E.~Tabanelli, D.~Brunelli, A.~Acquaviva, L.~Benini, Trimming feature extraction
  and inference for mcu-based edge nilm: A systematic approach, IEEE
  Transactions on Industrial Informatics 18~(2) (2021) 943--952.

\bibitem{iqbal2021critical}
H.~K. Iqbal, F.~H. Malik, A.~Muhammad, M.~A. Qureshi, M.~N. Abbasi, A.~R.
  Chishti, A critical review of state-of-the-art non-intrusive load monitoring
  datasets, Electric Power Systems Research 192 (2021) 106921.

\bibitem{klemenjak2019electricity}
C.~Klemenjak, A.~Reinhardt, L.~Pereira, S.~Makonin, M.~Berg{\'e}s,
  W.~Elmenreich, Electricity consumption data sets: Pitfalls and opportunities,
  in: Proceedings of the 6th ACM international conference on systems for
  energy-efficient buildings, cities, and transportation, 2019, pp. 159--162.

\bibitem{wu2020nonintrusive}
X.~Wu, D.~Jiao, L.~You, Nonintrusive on-site load-monitoring method with
  self-adaption, International Journal of Electrical Power \& Energy Systems
  119 (2020) 105934.

\bibitem{chavan2022iedl}
D.~R. Chavan, D.~S. More, A.~M. Khot, Iedl: Indian energy dataset with low
  frequency for nilm, Energy Reports 8 (2022) 701--709.

\bibitem{da2021deepdfml}
L.~da~Silva~Nolasco, A.~E. Lazzaretti, B.~M. Mulinari, Deepdfml-nilm: A new
  cnn-based architecture for detection, feature extraction and multi-label
  classification in nilm signals, IEEE sensors journal 22~(1) (2021) 501--509.

\bibitem{kelly2015uk}
J.~Kelly, W.~Knottenbelt, The uk-dale dataset, domestic appliance-level
  electricity demand and whole-house demand from five uk homes, Scientific data
  2~(1) (2015) 1--14.

\bibitem{Kriechbaumer2018BLONDAB}
T.~Kriechbaumer, H.-A. Jacobsen, Blond, a building-level office environment
  dataset of typical electrical appliances, Scientific data 5~(1) (2018) 1--14.

\bibitem{inproceedings}
M.~Kahl, A.~U. Haq, T.~Kriechbaumer, H.-A. Jacobsen, Whited-a worldwide
  household and industry transient energy data set, in: 3rd international
  workshop on non-intrusive load monitoring, 2016, pp. 1--4.

\bibitem{picon2016cooll}
T.~Picon, M.~N. Meziane, P.~Ravier, G.~Lamarque, C.~Novello, J.-C.~L. Bunetel,
  Y.~Raingeaud, Cooll: Controlled on/off loads library, a public dataset of
  high-sampled electrical signals for appliance identification, arXiv preprint
  arXiv:1611.05803 (2016).

\bibitem{plaid}
J.~Gao, S.~Giri, E.~C. Kara, M.~Berg{\'e}s, Plaid: a public dataset of
  high-resoultion electrical appliance measurements for load identification
  research: demo abstract, in: proceedings of the 1st ACM Conference on
  Embedded Systems for Energy-Efficient Buildings, 2014, pp. 198--199.

\bibitem{filip2011blued}
K.~Anderson, A.~Ocneanu, D.~Benitez, D.~Carlson, A.~Rowe, M.~Berges, Blued: A
  fully labeled public dataset for event-based non-intrusive load monitoring
  research, in: Proceedings of the 2nd KDD workshop on data mining applications
  in sustainability (SustKDD), Vol.~7, ACM New York, 2012, pp. 1--5.

\bibitem{kolter2011redd}
J.~Z. Kolter, M.~J. Johnson, Redd: A public data set for energy disaggregation
  research, in: Workshop on data mining applications in sustainability
  (SIGKDD), San Diego, CA, Vol.~25, Citeseer, 2011, pp. 59--62.

\bibitem{papageorgiou2025nilm}
P.~G. Papageorgiou, G.~C. Christoforidis, A.~S. Bouhouras, Nilm in high
  frequency domain: A critical review on recent trends and practical
  challenges, Renewable and Sustainable Energy Reviews 213 (2025) 115497.

\bibitem{silva2024review}
M.~D. Silva, Q.~Liu, A review of nilm applications with machine learning
  approaches., Computers, Materials \& Continua 79~(2) (2024).

\bibitem{renaux2020dataset}
D.~P.~B. Renaux, F.~Pottker, H.~C. Ancelmo, A.~E. Lazzaretti, C.~R.~E. Lima,
  R.~R. Linhares, E.~Oroski, L.~d.~S. Nolasco, L.~T. Lima, B.~M. Mulinari,
  et~al., A dataset for non-intrusive load monitoring: Design and
  implementation, Energies 13~(20) (2020) 5371.

\bibitem{kahl2019measurement}
M.~Kahl, V.~Krause, R.~Hackenberg, A.~Ul~Haq, A.~Horn, H.-A. Jacobsen,
  T.~Kriechbaumer, M.~Petzenhauser, M.~Shamonin, A.~Udalzow, Measurement system
  and dataset for in-depth analysis of appliance energy consumption in
  industrial environment, tm-Technisches Messen 86~(1) (2019) 1--13.

\bibitem{nour2023data}
M.~Nour, J.-C. Le~Bunetel, P.~Ravier, Y.~Raingeaud, Data augmentation
  strategies for high-frequency nilm datasets, IEEE Transactions on
  Instrumentation and Measurement 72 (2023) 1--9.

\bibitem{may2010data}
R.~J. May, H.~R. Maier, G.~C. Dandy, Data splitting for artificial neural
  networks using som-based stratified sampling, Neural Networks 23~(2) (2010)
  283--294.

\bibitem{shin2019data}
C.~Shin, S.~Rho, H.~Lee, W.~Rhee, Data requirements for applying machine
  learning to energy disaggregation, Energies 12~(9) (2019) 1696.

\bibitem{geng2025diffusion}
Z.~Geng, L.~Yang, W.~Yu, A diffusion model-based framework to enhance the
  robustness of non-intrusive load disaggregation, Energy 320 (2025) 135423.

\bibitem{goyal2024systematic}
M.~Goyal, Q.~H. Mahmoud, A systematic review of synthetic data generation
  techniques using generative {AI}, Electronics 13~(17) (2024) 3509.

\bibitem{7778841}
D.~Chen, D.~Irwin, P.~Shenoy, Smartsim: A device-accurate smart home simulator
  for energy analytics, in: 2016 IEEE International Conference on Smart Grid
  Communications (SmartGridComm), IEEE, 2016, pp. 686--692.

\bibitem{8340657}
N.~Buneeva, A.~Reinhardt, Ambal: Realistic load signature generation for load
  disaggregation performance evaluation, in: 2017 ieee international conference
  on smart grid communications (smartgridcomm), IEEE, 2017, pp. 443--448.

\bibitem{antgen}
A.~Reinhardt, C.~Klemenjak, How does load disaggregation performance depend on
  data characteristics? insights from a benchmarking study, in: Proceedings of
  the eleventh ACM international conference on future energy systems, 2020, pp.
  167--177.

\bibitem{SynD}
C.~Klemenjak, C.~Kovatsch, M.~Herold, W.~Elmenreich, A synthetic energy dataset
  for non-intrusive load monitoring in households, Scientific data 7~(1) (2020)
  108.

\bibitem{donnal2022nilm}
J.~Donnal, Nilm-synth: Synthetic dataset generation for non-intrusive load
  monitoring algorithms, in: 2022 IEEE Power \& Energy Society Innovative Smart
  Grid Technologies Conference (ISGT), IEEE, 2022, pp. 1--6.

\bibitem{li2022energy}
J.~Li, Z.~Chen, L.~Cheng, X.~Liu, Energy data generation with wasserstein deep
  convolutional generative adversarial networks, Energy 257 (2022) 124694.

\bibitem{HENRIET2018268}
S.~Henriet, U.~{\c{S}}im{\c{s}}ekli, B.~Fuentes, G.~Richard, A generative model
  for non-intrusive load monitoring in commercial buildings, Energy and
  Buildings 177 (2018) 268--278.

\bibitem{weisshaar2020expansion}
D.~Wei{\ss}haar, P.~Held, D.~O. Abdeslam, D.~Benyoucef, Expansion and
  superposition of switching cycles to generate simulation datasets for nilm,
  in: IECON 2020 The 46th Annual Conference of the IEEE Industrial Electronics
  Society, IEEE, 2020, pp. 5163--5169.

\bibitem{held2019generation}
P.~Held, D.~Wei{\ss}haar, D.~O. Abdeslam, D.~Benyoucef, Generation of new
  simulation scenarios for nilm based on real data sets using high-resolution
  current waveforms, in: IECON 2019-45th Annual Conference of the IEEE
  Industrial Electronics Society, Vol.~1, IEEE, 2019, pp. 5319--5324.

\bibitem{kamyshev2023physics}
I.~Kamyshev, S.~M. Hoosh, H.~Ouerdane, Physics-informed appliance signatures
  generator for energy disaggregation, in: 2023 IEEE 7th Conference on Energy
  Internet and Energy System Integration (EI2), IEEE, 2023, pp. 3591--3596.

\bibitem{hernandez2022synthetic}
M.~Hernandez, G.~Epelde, A.~Alberdi, R.~Cilla, D.~Rankin, Synthetic data
  generation for tabular health records: A systematic review, Neurocomputing
  493 (2022) 28--45.

\bibitem{iwana2021time}
B.~K. Iwana, S.~Uchida, Time series data augmentation for neural networks by
  time warping with a discriminative teacher, in: 2020 25th International
  Conference on Pattern Recognition (ICPR), IEEE, 2021, pp. 3558--3565.

\bibitem{liu2024non}
Y.~Liu, Y.~Wang, J.~Ma, Non-intrusive load monitoring in smart grids: A
  comprehensive review, arXiv preprint arXiv:2403.06474 (2024).

\bibitem{tracegan}
A.~Harell, R.~Jones, S.~Makonin, I.~V. Baji{\'c}, Tracegan: Synthesizing
  appliance power signatures using generative adversarial networks, IEEE
  Transactions on Smart Grid 12~(5) (2021) 4553--4563.

\bibitem{xiao2025non}
Y.~Xiao, Z.~Tan, B.~Qian, J.~Zhang, Y.~Luo, F.~Zhang, J.~Huang, X.~Feng, A
  non-intrusive load monitoring data generation method, in: Journal of Physics:
  Conference Series, Vol. 2963, IOP Publishing, 2025, p. 012016.

\bibitem{guyon2003design}
I.~Guyon, Design of experiments of the nips 2003 variable selection benchmark,
  in: NIPS 2003 workshop on feature extraction and feature selection, Vol. 253,
  2003, p.~40.

\bibitem{held2018frequency}
P.~Held, S.~Mauch, A.~Saleh, D.~O. Abdeslam, D.~Benyoucef, Frequency invariant
  transformation of periodic signals (fit-ps) for classification in nilm, IEEE
  Transactions on Smart Grid 10~(5) (2018) 5556--5563.

\bibitem{piasg}
I.~Kamyshev, S.~M. Hoosh, H.~Ouerdane, Physics-informed appliance signatures
  generator for energy disaggregation, in: 2023 IEEE 7th Conference on Energy
  Internet and Energy System Integration (EI2), IEEE, 2023, pp. 3591--3596.

\bibitem{alaa2022faithful}
A.~Alaa, B.~Van~Breugel, E.~S. Saveliev, M.~van~der Schaar, How faithful is
  your synthetic data? sample-level metrics for evaluating and auditing
  generative models, in: International Conference on Machine Learning, PMLR,
  2022, pp. 290--306.

\bibitem{flach2015precision}
P.~Flach, M.~Kull, Precision-recall-gain curves: Pr analysis done right,
  Advances in neural information processing systems 28 (2015).

\bibitem{features}
M.~Kahl, A.~Ul~Haq, T.~Kriechbaumer, H.-A. Jacobsen, A comprehensive feature
  study for appliance recognition on high frequency energy data, in:
  Proceedings of the Eighth International Conference on Future Energy Systems,
  2017, pp. 121--131.

\end{thebibliography}

\end{document}